\documentclass[%
 reprint,
amsmath,amssymb,
aps,
]{revtex4-1}

\usepackage{graphicx}
\usepackage{dcolumn}
\usepackage{bm}
\usepackage[colorlinks, linkcolor=blue, anchorcolor=blue, citecolor=blue]{hyperref}
\usepackage{url}
\usepackage{amsmath,amssymb}
\usepackage[mathlines]{lineno}
\usepackage{subfigure}
\usepackage{eurosym}
\usepackage{amsmath,amssymb,mathrsfs}

\begin{document}

\preprint{APS/123-QED} 

\title{Single photon scattering from a chain of giant atoms coupled to a one-dimensional waveguide}
\author{Y. P. Peng}
\affiliation{School of Physical Science and Technology, Southwest Jiaotong University, Chengdu 610031, China}
\author{W. Z. Jia}
\email{wenzjia@swjtu.edu.cn}
\affiliation{School of Physical Science and Technology, Southwest Jiaotong University, Chengdu 610031, China}

\date{\today}

\begin{abstract}
We investigate coherent single-photon transport in a waveguide quantum electrodynamics structure containing multiple giant atoms. 
The single-photon scattering amplitudes are solved using a real-space method. The results give rise to a clear picture of the multi-channel scattering process. In the case of identical and equally-spaced giant atoms in a separate configuration, we also use the transfer-matrix method  to express the scattering amplitudes in terms of compact analytical expressions, which allow us to conveniently analyze the properties of the scattering spectra. 
Based on these theoretical results, we find that the non-dipole effects of giant atoms, which are relevant to the design of the setup, can strongly manipulate several types of collective properties of the output fields, including the superradiant phenomenon, the multiple Fano interference, and the photonic band gap. This makes it possible to manipulate the photon transport in a more versatile way than with small atoms.
We also make a proposal to probe the topological states of a chain of braided giant atoms by using photon scattering spectra, showing that waveguide quantum electrodynamics systems with giant atoms are ideal platforms to merge topological physics and on-chip quantum optics.

\end{abstract}

\maketitle


\section{\label{introduction}INTRODUCTION}

Waveguide quantum electrodynamics (wQED) systems are realized by strongly coupling a single or multiple quantum emitters to a one-dimensional (1D) waveguide and can be used to manipulate light-matter interactions at the single-photon level \cite{Roy-RMP2017, Gu-PhysReports2017,Sheremet-RMP2023}. 
The high atom-waveguide coupling efficiency makes these systems excellent platforms for controlling the transport of single or few photons \cite{Shen-OL2005,Shen-PRL2005,Chang-PRL2006,Shen-PRL2007,Zhou-PRL2008,Shi-PRB2009,Astafiev-Science2010,Longo-PRL2010,Zheng-PRA2010,Fan-PRA2010,Roy-PRL2011,Zheng-PRL2011,Fang-PRA2015,Shi-PRA2015}, and may have potential applications in quantum devices at the single-photon level \cite{Chang-Natphy2007,Aoki-PRL2009,Abdumalikov-PRL2010,Hoi-PRL2011,Bradford-PRL2012,Hoi-PRL2013,Neumeier-PRL2013,Jia-PRA2017,Zhu-PRA2017}. In particular, when multiple emitters are coupled to a 1D waveguide, the coherent and dissipative interactions between the emitters can be mediated by the propagation modes, leading to many interesting phenomena, including superradiant and subradiant states \cite{PR-Dicke1954,Vetter-PScr2016,Loo-Science2013,Zhang-PRL2019,Ke-PRL2019,Wang-PRL2020,Dinc-PRR2019,Dinc-Quantum2019}, 
long-range entanglement between distant emitters \cite{Zheng-PRL2013,Ballestero-PRA2014,Facchi-PRA-2016,Mirza-PRA2016},  
photonic bandgap generation \cite{Fang-PRA2015,Greenberg-PRA2021,He-OL2021},   
cavity-QED with atomic mirrors \cite{Chang-NJP2012,Mirhosseini-Natrue2019}, topologically protected spectroscopy \cite{Nie-PRR2020} and
topology-enhanced nonreciprocal scattering \cite{Nie-PRApplied2021}, 
multiple Fano interferences and  electromagnetically-induced-transparency-(EIT-)like phenomena \cite{Tsoi-PRA2008,Cheng-OL2012,Liao-PRA2015,Cheng-PRA2017,Ruostekoski-PRA2017,Mukhopadhyay-PRA2019,Mukhopadhyay-PRA2020,Jia-EPJP2022}, and so on. 

Recently, wQED systems with giant atoms \cite{Kockum-MI2021,Kockum-PRA2014} represent a new paradigm in quantum optics, in which the quantum systems, such as superconducting artificial atoms \cite{Kannan-Nature2020}, cold atoms in optical lattices \cite{Tudela-PRL2019}, giant spin ensembles \cite{Wang-Natcom2023}, etc., are simultaneously coupled to distant positions (with wavelength spacings) of the photonic bath. Thus, the giant atoms cannot be considered as point-like particles, and the usual dipole approximation is no longer valid. The non-dipole effects of giant-atom systems can produce some novel phenomena, such as frequency-dependent decay rate and Lamb shift \cite{Kockum-PRA2014,Kannan-Nature2020,Vadiraj-PRA2021}, decoherence-free interaction between braided giant atoms \cite{Kannan-Nature2020,Kockum-PRL2018,Carollo-PRR2020,Du-PRA2023}, non-Markovian dynamics \cite{Guo-PRA2017,Andersson-NatPhy2019}, generation of enhanced entanglement \cite{Santos-PRL2023}, tunable atom-photon bound states \cite{Guo-PRR2020,Guo-PRA2020,Zhao-PRA2020,Wang-PRL2021,Cheng-PRA2022,Soro-PRA2023,Jia-ArXiv2023} and scattering states \cite{Guo-PRA2017,Ask-ArXiv2020,Cai-PRA2021,Feng-PRA2021,Zhu-PRA2022,Yin-PRA2022}, and so on. 
The giant-atom effects on chiral atom-waveguide coupling \cite{Soro-PRA2022} and ultra strong coupling \cite{Noachtar-PRA2022,Sergi-PRA2022} have also been investigated. The structure of giant atom can also be implemented in a synthetic dimension \cite{Du-PRL2022,Xiao-npj2022}. 

The study of scattering spectra is an important aspect of wQED systems. On one hand, the light-matter interactions during the 
scattering process make the photons be effectively controlled. On the other hand, photon scattering spectra can reveal  useful 
information about the interactions between light and matter. It is shown that in wQED structures containing a single giant atom, effects 
beyond the dipole approximation play an important role in photon transport \cite{Guo-PRA2017,Cai-PRA2021}. For the double giant atom 
structure, the coherent and dissipative interactions between atoms, which can be manipulated by designing the layout of the coupling 
points, can produce more abundant scattering spectra than small atoms \cite{Ask-ArXiv2020,Feng-PRA2021,Yin-PRA2022}.
In this paper, we focus on the most general case of photon scattering from a collection of giant atoms, each with multiple coupling points, 
coupled to a linear waveguide. Using the real-space 
method, we obtain the expressions for the atomic excitation amplitudes and the scattering coefficients. A clear physical 
picture of the scattering process from the point of view of collective excitation is given. We also establish a mapping between an array of 
\textit{separate} giant atoms and a chain of small atoms. Based on this, we generalize the transfer-matrix approach for small atoms 
\cite{Tsoi-PRA2008,Mukhopadhyay-PRA2019,Mukhopadhyay-PRA2020,Jia-EPJP2022} to the case of separate giant atoms. 
Using the above theoretical tools, we analyze in detail the characteristics of the scattering spectra for multi-giant-atom systems in different configurations.
In particular, for an array of identical and periodically arranged atoms in a separate configuration, by analyzing the explicit analytical expressions under different parameters, we find that the non-dipole effects of giant atoms can strongly manipulate several types of collective properties of 
the output fields, including the superradiant phenomenon, the multiple Fano interference, and the photonic band gap. This allows for more versatile manipulation of photon transport than an array of small atoms. The formation of these spectral structures is well interpreted as interferences between different scattering channels relevant to the excitation of the collective modes. 
For an array of braided giant atoms, we focus on the influence of the unique decoherence-free interactions on the scattering spectra.
By designing the layout of the connecting points and the corresponding decoherence-free interactions, we construct an effective atomic array described by the Su-Schrieffer-Heeger (SSH) model \cite{Su-PRL1979}, and further make a proposal to probe the topological states of a braided atomic chain. It is shown that the photon scattering spectra can provide useful informations of the interactions between light and unconventional many-body states, and establish a bridge between topological physics and on-chip quantum optics. Conversely, 
the systems can be used to realize tunable and topologically protected photon transport. 

The remainder of this paper is organized as follows. In Sec.~\ref{Model}, we introduce the basic model of the wQED structure containing multiple giant atoms and further obtain the most general expressions of single-photon scattering amplitudes, which are applicable for arbitrary configuration. In Sec.~\ref{SpectraSeparated}, based on explicit analytical expressions obtained from the transfer-matrix approach, we provide a detailed analysis on the scattering spectra for an array of separate giant atoms. In Sec.~\ref{BraidedDFI}, we analyze the spectra that can reveal non-trivial many-body states caused by decoherence-free interactions in an array of braided giant atoms.
Finally, further discussions and conclusions are given in Sec.~\ref{conclusion}.
\section{\label{Model}MODEL AND Solution}

\subsection{\label{Hamiltonian}Hamiltonian and equations of motion}
\begin{figure}[t]
	\centering
	\includegraphics[width=0.5\textwidth]{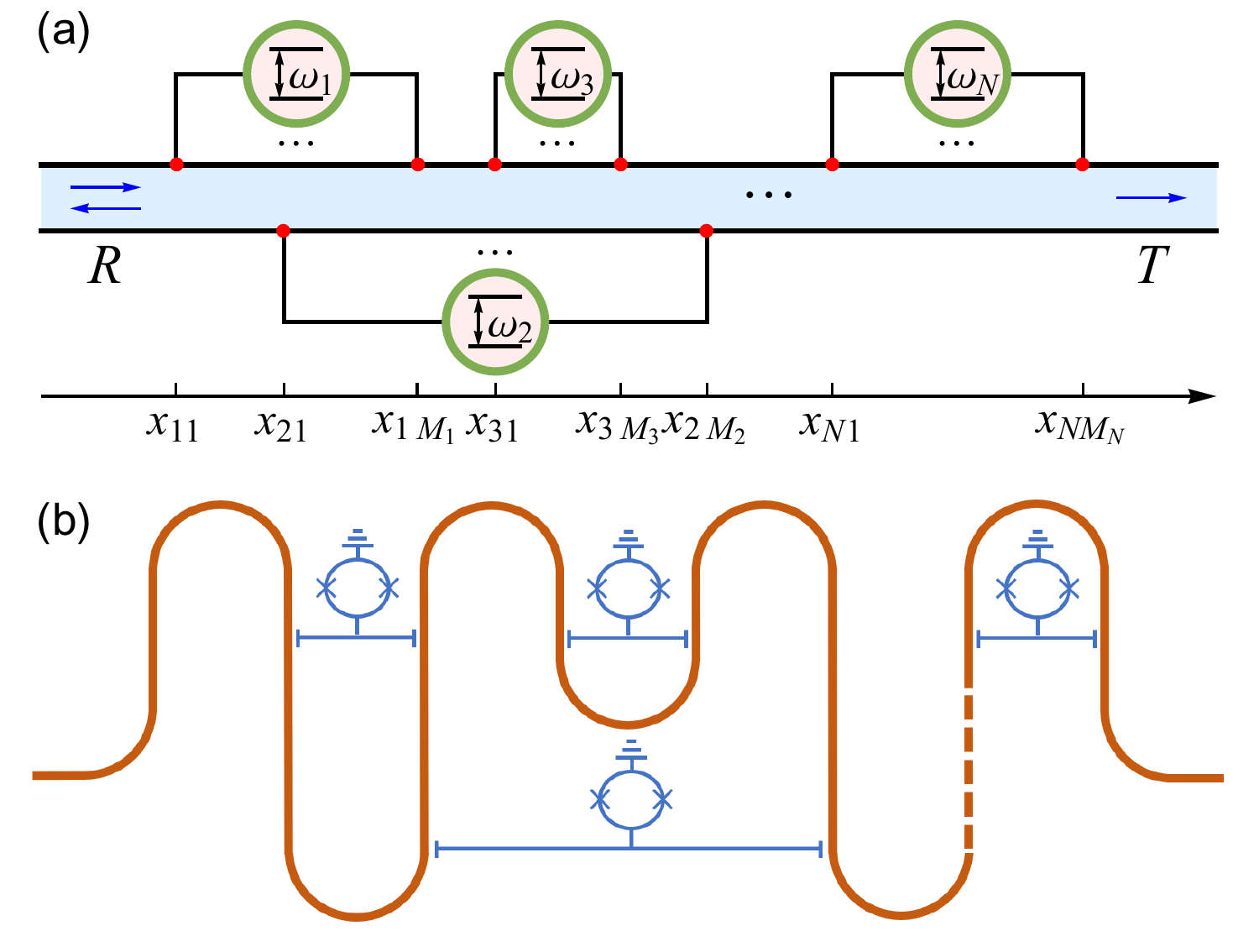}
	\caption{(a) A schematic of multiple two-level giant atoms coupled to a 1D waveguide through multiple connection points. (b) 
	A possible implementation with superconducting circuits, where multiple Xmon qubits coupled capacitively to a meandering microwave  
	transmission line at multiple points.
	}
	\label{SketcheGeneral}
\end{figure}
Here we focus on the wQED structures with multiple two-level
giant atoms, where each atom is coupled to a 1D waveguide
through multiple connection points, as shown in Fig.~\ref{SketcheGeneral}(a). In Fig.~\ref{SketcheGeneral}(b), we show how this setup could be implemented with superconducting qubits coupled to a meandering coplanar waveguide. 
We assume that the total number of atoms is $N$, and the number of coupling points for the $i$th atom is $M_i$. Under the
rotating-wave approximation, the Hamiltonian of the
system in the real space \cite{Roy-RMP2017} can be written as ($\hbar=1$)
\begin{equation}
	\begin{aligned}
	 \hat{H}=&\sum_{i=1}^{N}{\omega_i\hat{\sigma}_{i}^{+}\!\:\hat{\sigma}_{i}^{-}}+\int\mathrm{d}x\sum_s{\hat{c}_{s}^{\dagger}(x)\left(-\mathrm{i}l_sv_{\mathrm{g}}\frac{\partial}{\partial x} \right)\hat{c}_s\left( x \right)}
	 \\
	&+\int\mathrm{d}x\sum_{s}\sum_{i=1}^{N}\sum_{m=1}^{M_i}{V_{im}}\delta \left( x-x_{im} \right) \left[ \hat{c}_{s}^{\dagger}\left( x \right) \hat{\sigma}_{i}^{-}+\mathrm{H.c.} \right],
\end{aligned}
\end{equation}
where $s=\mathrm R, \mathrm L$, $l_{\mathrm R,\mathrm L} = \pm1$. $\hat{\sigma}_{i}^{+}$ ($\hat{\sigma}_{i}^{-}$)
is the raising (lowering) operator of
the atom $i$. $\hat{c}_{R}^{\dagger}(x)$ [$\hat{c}_{R}(x)$] and
 $\hat{c}_{L}^{\dagger}(x)$ [$\hat{c}_{L}(x)$] are the field operators of creating (annihilating)
the right- and left-propagating photons at position $ x$ in the
waveguide. $\omega_i$ is the 
transition frequency of the atom $i$. $v_{\mathrm{g}}$ is the group velocity of the photons in the
waveguide. $V_{im}$ is the
coupling strength of the $m$th coupling point of the $i$th giant atom at position $x_{im}$.

We assume that initially a single photon with energy $\omega =
v_{\mathrm{g}} k$ is incident from the left, where $k$ is the wave vector of
the photon. In the single-excitation subspace, the interacting
eigenstate of the system can be written as
\begin{equation}
|\Psi \rangle =\sum_s{\int{d}}x\Phi_s(x)\hat{c}_{s}^{\dagger}(x)|\emptyset \rangle +\sum_{i=1}^{N}{f_i}\hat{\sigma}_{i}^{+}|\emptyset\rangle,
\label{IntEigenStat}
\end{equation}
where $|\emptyset\rangle$ is the vacuum state, which means that there are no photons in the waveguide, and meanwhile the atoms are in their ground states. $\Phi_{s}(x)$ ($s=\mathrm{R}, \mathrm{L}$) is the single-photon wave function in the $s$ mode. $f_{i}$ is the excitation amplitude of the atom $i$. Substituting Eq.~\eqref{IntEigenStat} into the eigenequation
\begin{equation}
\hat{H}|\Psi\rangle=\omega|\Psi\rangle
\end{equation}
yields the following equations of motion:
\begin{subequations}
\begin{equation}
\left(-\mathrm{i} v_{\mathrm{g}} \frac{\partial}{\partial x}-\omega\right) \Phi_\mathrm{R}(x)+\sum_{i=1}^{N}\sum_{m=1}^{M_i}V_{i m}\delta\left(x-x_{im}\right) f_{i}=0,
\label{EoM1}
\end{equation}	
\begin{equation}
\left(\mathrm{i} v_{\mathrm{g}} \frac{\partial}{\partial x}-\omega\right) \Phi_\mathrm{L}(x)+\sum_{i=1}^{N}\sum_{m=1}^{M_i} V_{i m} \delta\left(x-x_{im}\right) f_{i}=0,
\label{EoM2}
\end{equation}
\begin{equation}
\left(\omega_{i}-\omega\right) f_{i}+\sum_{s}\sum_{m=1}^{M_i} V_{i m}\Phi_{s}\left(x_{i m}\right)=0.
\label{EoM3}
\end{equation}
\end{subequations}
\begin{widetext}
\subsection{\label{GeneralExpression}General expressions for scattering amplitudes}
For a photon incident from the left, $\Phi_\mathrm{R}(x)$ and $\Phi_\mathrm{L}(x)$ take the following ansatz
\begin{subequations}
	\begin{equation}
\Phi_\mathrm{R}(x)=e^{\mathrm{i}kx}\left[ \vartheta \left( x_1-x \right) +\sum_{p=1}^{N_\mathrm{c}-1}{t_p}
\vartheta 
\left( x-x_p \right) \vartheta \left( x_{p+1}-x \right) +t\vartheta \left( x-x_{N_\mathrm{c}}\right) \right], 
\label{PhiR}
	\end{equation}	
	\begin{equation}
\Phi_\mathrm{L}(x)=e^{-\mathrm{i}kx}\left[ r\vartheta \left( x_1-x \right) +\sum_{p=2}^{N_\mathrm{c}}{r_p}\vartheta \left( x-x_{p-1} \right) \vartheta \left( x_p-x \right) \right]. 
\label{PhiL}
	\end{equation}
\end{subequations}
Here the positions of coupling points from left to right are labeled as $x_p$. $N_\mathrm{c}$ is the total number of  
coupling points and satisfies the relation $\sum_i^N M_i=N_\mathrm{c}$. Here $t_p$ ($r_p$) is the transmission (reflection) 
amplitude of the $p$th coupling point, $t$ ($r$) is the transmission (reflection) amplitude
of the last (first) coupling point, and $\vartheta (x)$ denotes the Heaviside step function.
\end{widetext}

Starting from the equations of motion \eqref{EoM1}-\eqref{EoM3} and the ansatz 
\eqref{PhiR}-\eqref{PhiL}, and after some algebra (see Appendix \ref{Derivation1} for details), we can obtain the atomic excitation amplitudes 
\begin{equation}
\mathbf{f}=\sqrt{v_{\mathrm{g}}}\left(\omega\mathbf{I}-\mathbf{H} \right)^{-1}\mathbf{V},
\label{fGeneral}
\end{equation}
and the corresponding transmission and reflection amplitudes
\begin{subequations}
\begin{equation}
t=1-\frac{\mathrm{i}}{\sqrt{v_{\mathrm{g}}}}\mathbf{V}^{\dag}\mathbf{f}=1-{\mathrm{i}}\mathbf{V}^{\dag}\left(\omega\mathbf{I}-\mathbf{H} \right)^{-1}\mathbf{V},
\label{tGeneral}
\end{equation}
\begin{equation}
r=-\frac{\mathrm{i}}{\sqrt{v_{\mathrm{g}}}}\mathbf{V}^{\top}\mathbf{f}=-{\mathrm{i}}\mathbf{V}^{\top}\left(\omega\mathbf{I}-\mathbf{H} \right)^{-1}\mathbf{V}.
\label{rGeneral}
\end{equation}
\end{subequations}
Here $\mathbf{I}$ is the identity matrix. $\mathbf{f}$ is written as 
\begin{equation}
\mathbf{f}=( f_1,f_2,\cdots f_{N}) ^{\top}. 
\label{fmatrix}
\end{equation}
$\mathbf{V}$ takes the form 
\begin{equation}
\mathbf{V}=(\mathcal{V}_1,\mathcal{V}_2,\cdots,\mathcal{V}_{N})^{\top},
\label{Vmatrix}
\end{equation}
with elements
\begin{equation} 
\mathcal{V}_{i}=\sum_{m=1}^{M_i}{\sqrt{\frac{\gamma _{im}}{2}}}e^{\mathrm{i}\theta_{im}(\omega)},
\end{equation}
where the phase factor is defined as $\theta_{im}(\omega)=kx_{im}=\omega x_{im}/v_{\mathrm{g}}$ and the decay rate into the guided modes through the coupling point at $x_{im}$ is $\gamma_{im}=2V_{im}^2/v_{\mathrm{g}}$ from Fermi’s golden rule. $\mathbf{H}$ is the effective non-Hermitian Hamilton matrix of the atom array, with elements 
\begin{equation}
\mathcal{H}_{ij}=\omega_i\delta _{ij}-\frac{\mathrm{i}}{2}\sum_{m=1}^{M_i}\sum_{m'=1}^{M_j}{{\sqrt{\gamma _{im}\gamma _{jm'}}}}e^{\mathrm{i}\left| \theta _{im}(\omega)-\theta _{jm'}(\omega) \right|}.
\label{EffH}
\end{equation}
The non-diagonal element of this effective non-Hermitian Hamilton matrix describes the coherent and dissipative atom-atom interactions mediated by the waveguide modes. The effective Hamiltonian of the atom array takes the form 
\begin{equation}
\hat H_{\mathrm{eff}}=\sum_{i=1}^{N}\sum_{j=1}^{N}\mathcal{H}_{ij}\hat\sigma^{+}_i\hat\sigma^{-}_j.
\label{Heff}
\end{equation}
Note that Eqs.~\eqref{fGeneral}-\eqref{rGeneral} are applicable for the most general setup possible, i.e.,
an arbitrary number of giant atoms with an arbitrary number of connection points each.

To better understand the physics of the scattering process, we rewrite the scattering amplitudes in terms of 
collective modes of the atoms 
\begin{subequations}
	\begin{equation}
		t =1-\mathrm{i}\sum_{n=1}^{N}\frac{\big(\mathbf{V}^{\dagger}{\mathbf{U}^{\mathrm{\mathscr{R}}}_{n}}\big)\big({\mathbf{U}^{\mathscr{L}}_{n}}^\dagger\mathbf{V}\big)}{\omega-\lambda_n},
		\label{tDecompose1}
	\end{equation}
	\begin{equation}
		r =-\mathrm{i}\sum_{n=1}^{N}\frac{\big(\mathbf{V}^{\top}{\mathbf{U}^{\mathscr{R}}_{n}}\big)\big({\mathbf{U}^{\mathscr{L}}_{n}}^\dagger\mathbf{V}\big)}{\omega-\lambda_n},
		\label{rDecompose1}
	\end{equation}	
\end{subequations}
where $\mathbf{U}^{\mathscr{R}}_{n}$ and $\mathbf{U}^{\mathscr{L}}_{n}$ are the right and left eigenvectors of the non-Hermitian Hamilton matrix $\mathbf{H}$, satisfying ${\mathbf{U}^{\mathscr{L}}_{n}}^\dag\mathbf{U}^{\mathscr{R}}_{n'}=\delta_{nn'}$ \cite{Brody-JPA2013}. $\lambda_n$ is the complex eigenvalue of $\mathbf{H}$.
The numerators in the above equations represent the overlap degree of the $n$th collective state of atoms and the propagating photon modes. The transmittance  and the reflectance can be further defined as $T = |t|^2$ and $R = |r|^2$. Equations \eqref{tDecompose1} and \eqref{rDecompose1} show that the scattering spectrum can be regarded as the result of interference between different scattering channels, which are provided by the corresponding collective modes.
It's important to note that the above results apply to both the Markovian and the non-Markovian regimes. 

In what follows, we assume that the frequencies of all the giant atoms are distributed in a small range around a reference frequency, with $\omega_i\simeq\omega_\mathrm{a}$.
We further assume that the spacing between the connection points is small enough so that the non-Markovian effects (the phase-accumulated effects for detuned photons) can be neglected, thus the phase factors
$\theta_{im}(\omega)$ can be replaced by $\theta_{im}(\omega_\mathrm{a})=\omega_\mathrm{a}  x_{im}/v_g\equiv\theta_{im}$. Under this approximation, the elements of the effective Hamilton matrix $\mathbf{H}$ can be written as 
\begin{subequations}
\begin{equation}
\mathcal{H}_{ii}=\omega_{i}+\Delta_{\mathrm{L},i}-\frac{\mathrm{i}}{2}\Gamma _{\mathrm{eff},i},
\label{EffHMarkovii}
\end{equation}
\begin{equation}
\mathcal{H}_{ij,i\neq j}=g_{ij}-\frac{\mathrm{i}}{2}\Gamma_{\mathrm{coll},ij},
\label{EffHMarkovij}
\end{equation}
\end{subequations}
where the Lamb shift and the effective decay of the $i$th atom are defined as \cite{Kockum-PRL2018}
\begin{subequations}
\begin{equation}
\Delta _{\mathrm{L},i}=\frac1{2}\sum_{m=1}^{M_i}\sum _{m'=1}^{M_i}\sqrt{\gamma _{im}\gamma _{im'}}\sin |\theta_{im}-\theta_{im'}|,
\label{Deltaeff}
\end{equation}	
\begin{equation}
\Gamma _{\mathrm{eff},i}=\sum_{m=1}^{M_i}\sum _{m'=1}^{M_i}\sqrt{\gamma _{im}\gamma _{im'}}\cos(\theta_{im}-
\theta_{im'}).
\label{Gammaeff}
\end{equation}	
\end{subequations}
The exchange interaction and the collective decay between the $i$th and the $j$th ($i\neq j$) atom take the form  
\begin{subequations}
\begin{equation}
g_{ij}=\frac{1}{2}\sum_{m}^{M_i}\sum_{m'}^{M_j}\sqrt{\gamma_{im}\gamma_{jm'}}\sin\left|\theta_{im}-\theta_{jm'}\right|,
\label{CoherentInt}
\end{equation}
\begin{equation}
\Gamma_{\mathrm{coll},ij}=\sum_{m}^{M_i}\sum_{m'}^{M_j}\sqrt{\gamma_{im}\gamma_{jm'}}\cos\left(\theta_{im}-\theta_{jm'}\right).
\label{CollDecay}
\end{equation}
\end{subequations}
Clearly, the scattering spectra are determined by the above characteristic quantities.

Since under the Markovian approximation, the quantities $\mathbf{V}$ and $\mathbf{H}$ (and the corresponding 
eigenvalues $\lambda_n$ and eigenvectors $\mathbf{U}^{\mathscr{R,L}}_{n}$) are independent of the frequency of the 
photons, the energy and the effective decay of the $n$th collective mode can be defined as $\tilde{\omega}_n=\mathrm{Re}[\lambda_n(\omega_\mathrm{a})]$ and $\tilde{\Gamma}_{n} = -2\mathrm{Im}[\lambda_n(\omega_\mathrm{a})]$, respectively. Further, we can define the numerators in Eqs.~\eqref{tDecompose1} and \eqref{rDecompose1} as $\eta_n =-\mathrm{i}[{\big(\mathbf{V}^{\dagger}{\mathbf{U}^{\mathscr{R}}_{n}}\big)\big({\mathbf{U}^{\mathscr{L}}_{n}}^\dagger\mathbf{V}\big)}]|_{\omega=\omega_\mathrm{a}}$ and $\tilde\eta_n=-\mathrm{i}[{\big(\mathbf{V}^{\top}{\mathbf{U}^{\mathscr{R}}_{n}}\big)\big({\mathbf{U}^{\mathscr{L}}_{n}}^\dagger\mathbf{V}\big)}]|_{\omega=\omega_\mathrm{a}}$. Thus the transmission and reflection amplitudes can be written as
\begin{subequations}
	\begin{equation}
		t =1+\sum_{n=1}^{N}\frac{\eta_n}{\Delta-\tilde{\delta}_{n}+\mathrm{i}\frac{\tilde{\Gamma}_{n}}{2} },
		\label{tDecompose2}
	\end{equation}	
	\begin{equation}
		r =\sum_{n=1}^{N}\frac{\tilde\eta_n}{\Delta-\tilde{\delta}_{n}+\mathrm{i}\frac{\tilde{\Gamma}_{n}}{2}}.
		\label{rDecompose2}
	\end{equation}	
\end{subequations}
Here $\Delta=\omega-\omega_\mathrm{a}$ ($\tilde{\delta}_n=\tilde{\omega}_n-\omega_\mathrm{a}$) is the detuning between the frequency of the photons (the $n$th collective mode) and the reference frequency. 
Equations \eqref{tDecompose2} and \eqref{rDecompose2} show superpositions of several Lorentzian-type amplitudes contributed by the collective excitations.
$\eta_n$ ($\tilde\eta_n$) determines the weight of each Lorentzian component. Equations \eqref{tDecompose2} and \eqref{rDecompose2} are very helpful for us to analyze the scattering spectra for setups containing multiple giant atoms.
\begin{figure*}[t]
\centering
\includegraphics[width=\textwidth]{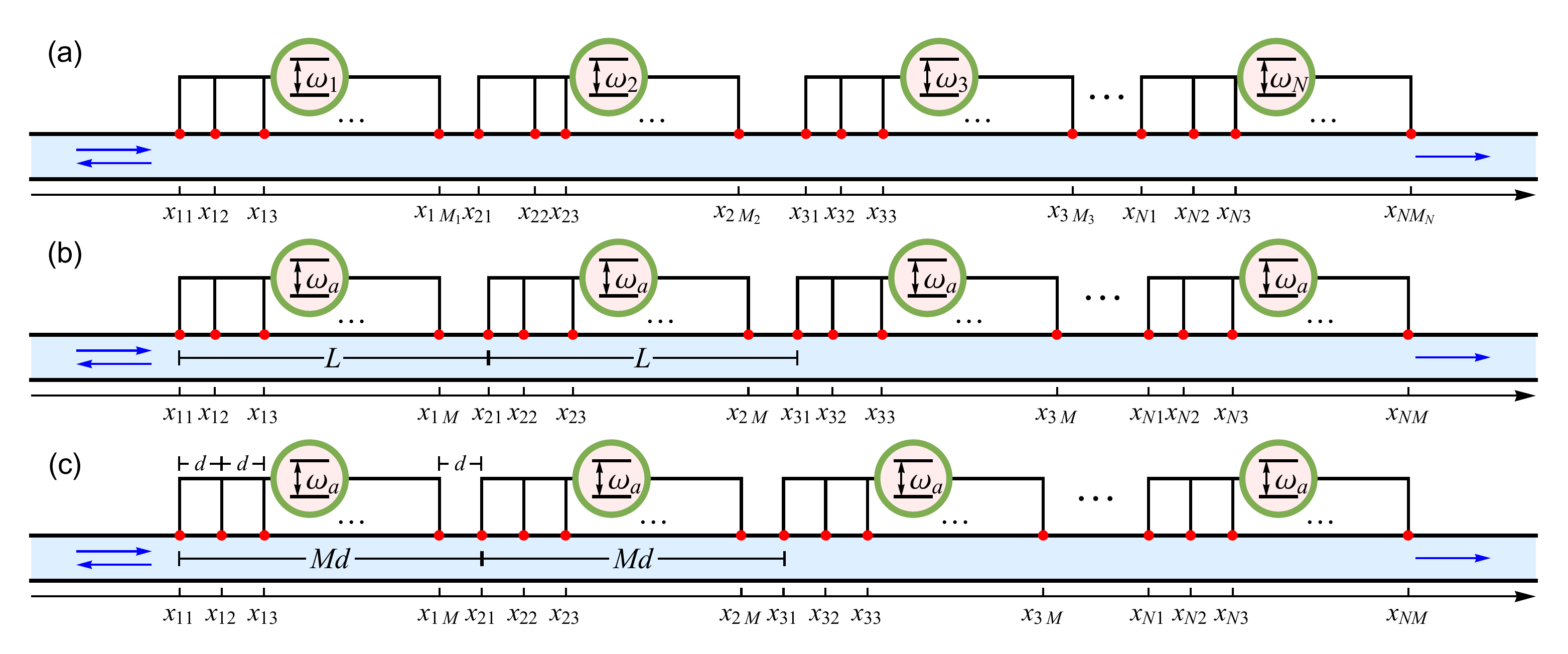}
\caption{\label{skec3}Sketches of 1D chains of separate giant atoms: (a) the most general configuration possible for an array of separate atoms, (b) 
the configuration where all the giant atoms are identical and periodically arranged, and (c) the configuration where all the atoms are identical and 
all the coupling points are equally spaced.}
\label{SeperateSketch}
\end{figure*}
\section{\label{SpectraSeparated}Spectra for an array of separate giant atoms}
The expressions \eqref{tDecompose1} and \eqref{rDecompose1} [or Eqs.~\eqref{tDecompose2} and \eqref{rDecompose2}] are applicable for any 
multi-emitter system with arbitrary coupling points, and reflect the physical picture of multi-channel scattering through the collective modes. 
However, for an array of separate giant atoms (see Fig.~\ref{SeperateSketch}), the photon scattering can be understood from a different point of 
view: the cascade scattering process (i.e., the output of the previous atom is the input of the next one). Thus we will show below that the 
transfer-matrix approach, which has been used to calculate the photon scattering amplitudes of an array of small atoms \cite{Tsoi-PRA2008,Mukhopadhyay-PRA2019,Mukhopadhyay-PRA2020,Jia-EPJP2022}, can be generalized to the case of separate giant atoms. In 
particular, if all the giant atoms are identical and periodically arranged,
we can obtain the explicit analytical expressions for the transmission and reflection amplitudes. 
By combining these results with those of the previous section, a number of properties of the scattering spectra can be determined and interpreted. 
\subsection{\label{GeneralTransferMatrix} Transfer matrix for arbitrary array of giant atoms in a separate configuration}
First, we consider an arbitrary giant atom array in the separate configuration, as shown schematically in 
Fig.~\ref{SeperateSketch}(a). We can prove that the scattering amplitudes on the left and right of the $i$th atom can be connected by the following recursive linear matrix equation (see Appendix \ref{DerivationTM} for details) 
\begin{equation}
\begin{pmatrix}t_{i-1,M_{i-1}}\\ r_{i1}\\ \end{pmatrix} =\mathbf{T}_{\alpha_i}^{-1}\mathbf{T}_{i}\mathbf{T}_{\alpha_i}\begin{pmatrix} t_{iM_{i}}\\ r_{i+1,1}\\ \end{pmatrix},
\label{recursiveM}
\end{equation}
with
\begin{subequations}
\begin{equation}
\mathbf{T}_i=\left(
\begin{array}{cc}
	1+\text{i}\xi_i & \text{i}\xi_i  \\
	-\text{i}\xi_i  & 1-\text{i}\xi_i \\
\end{array}
\right),
\label{Tmatrix}
\end{equation}	
\begin{equation}
\mathbf{T}_{\alpha_i}=\left( \begin{matrix}
	e^{\text{i}\alpha _i}&		0\\
	0&		e^{-\text{i}\alpha _i}\\
\end{matrix} \right).
\end{equation}
\end{subequations}
Here $\xi_i$ takes the form
\begin{equation}
	\xi_i=\frac{\Gamma _{\mathrm{eff},i}}{2 \left(\Delta _{i}-\Delta _{\mathrm{L},i}\right)},
	\label{xi}
\end{equation}	 
where $\Delta_i=\omega-\omega_i$ is the detuning between the photons and the $i$th atom. The Lamb shift $\Delta _{\mathrm{L},i}$ and the effective decay $\Gamma _{\mathrm{eff},i}$ of the $i$th atom are defined in Eqs.~\eqref{Deltaeff} and \eqref{Gammaeff}. 
The phase factor $\alpha_i$ is relevant to the effective position of the $i$th atom and satisfies
\begin{equation}
\tan2\alpha_i=\frac{\sum _{m,m'=1}^{M_i}\sqrt{\gamma _{im}\gamma _{im'}}\sin \left(\theta_{im}+\theta_{im'}\right)}{\sum _{m,m'=1}^{M_i}\sqrt{\gamma _{im}\gamma _{im'}}\cos \left(\theta_{im}+\theta_{im'}\right)}.
\label{tanalpha}
\end{equation}
The boundary conditions satisfied by Eq.~\eqref{recursiveM} are $t_{NM_N}=t$, $r_{N+1,1}=0$, $t_{0,M_0}=1$ and $r_{11}=r$. Note that here we have assumed that the transition frequencies of all the atoms are approximately equal $\omega_i\simeq\omega_\mathrm{a}$ and have used the Markovian approximation (i.e., ignore the phase-accumulated
effects for detuned photons). Thus, similar to Sec.~\ref{GeneralExpression}, the phase factor is defined as $\theta_{im}=\omega_\mathrm{a}x_{im}/v_g$, i.e., the wave vector $k$ has been replaced by $\omega_\mathrm{a}/v_g$.

Using Eq.~\eqref{recursiveM} iteratively $N$ times in succession and setting $\alpha_1=0$ (This can be achieved by selecting the origin of the coordinates appropriately), we obtain the following connective relation between the reflection and transmission amplitudes:
	\begin{equation}
\left( \begin{array}{c}
	1\\
	r\\
\end{array} \right) =\mathbf{M}\left( \begin{array}{c}
	t e^{\mathrm{i}\alpha_N}\\
	0\\
\end{array} \right) ,
	\label{simultaneous equation1}
\end{equation}
with
\begin{equation}
\mathbf{M}=
\left[\prod_{i=1}^{N-1}{\left(\mathbf{T}_i\mathbf{T}_{\phi_i} \right)}\right]\mathbf{T}_N.
\end{equation}
Here the effective phase propagating matrix is defined as
\begin{equation}
\mathbf{T}_{\phi_i}=\mathbf{T}_{\alpha_i}\mathbf{T}^{-1}_{\alpha_{i+1}}=\left(
   	\begin{array}{cc}
   		e^{-\text{i$\phi $}_i} & 0 \\
   		0 & e^{\text{i$\phi $}_i} \\
   	\end{array}
   	\right),
	\label{Tphi_i}
\end{equation}
with $\phi_i=\alpha_{i+1}-\alpha_{i}$. Straightforwardly, the transmission and reflection amplitudes read
\begin{subequations}
	\begin{equation}
t=\frac{e^{-\text{i}\alpha_N}}{\mathbf{M}_{11}},
		\label{finalt}
	\end{equation}	
	\begin{equation}
r=\frac{\mathbf{M}_{21}}{\mathbf{M} _{11}},
		\label{finalr}
	\end{equation}
\end{subequations}
which are applicable to any giant-atom array in separate configuration.
In fact, the above results indicate that $N$ separate giant atoms in a wQED structure are equivalent to $N$ small atoms, each with a frequency shift $\Delta_{\mathrm{L},i}$, an effective decay $\Gamma_{\mathrm{eff},i}$, and a reference phase $\alpha_{i}$ (relevant to the effective position of the $i$th atoms).
\subsection{\label{GeneralTransferMatrix} Transfer matrix for an array of periodically arranged identical giant atoms in a separated configuration}
Here we focus on the case where all the giant atoms are identical and periodically arranged, as shown schematically in Fig.~\ref{SeperateSketch}(b). This require the following conditions: 
(i) All transition frequencies are equal, with $\omega_{i}=\omega_{\mathrm{a}}$, i.e., all detunings are equal, with $\Delta_i=\omega-\omega_{\mathrm{a}}\equiv\Delta$. 
(ii) The number and layout of the connection points are the same for each atom, with $M_i=M$, $\theta_{im}-\theta_{in}=\theta_{jm}-\theta_{jn}$, and $\gamma_{im}=\gamma_{jm}$. 
(iii) The atoms are evenly spaced, with lattice constant $x_{i+1,m}-x_{im}= L$, i.e., the phase factors satisfy the relation $\theta_{i+1,m}-\theta_{im}=\omega_\mathrm{a} L/v_\mathrm{g}\equiv\phi$. 

Thus all the Lamb shifts and the effective decays are identical, with $\Delta_{\mathrm{L},i}=\Delta_{\mathrm{L}}$ and $\Gamma_{\mathrm{eff},i}=\Gamma _{\mathrm{eff}}$. In addition, from Eq.~\eqref{tanalpha}, one can prove that for this case $\phi_i=\alpha_{i+1}-\alpha_{i}=\phi$, i.e., $\alpha_{i}=(i-1)\phi$, is satisfied. The connective relation Eq.~\eqref{simultaneous equation1} becomes
\begin{equation}
	\left( \begin{array}{c}
		1\\
		r\\
	\end{array} \right) ={{\tilde{\mathbf{T}}^N}}\left( \begin{array}{c}
		te^{\text{i}N\phi}\\
		0\\
	\end{array} \right),
	\label{simultaneous equation3}
\end{equation}
with
\begin{equation}
\tilde{\mathbf{T}}=\left(
\begin{array}{cc}
	(1+\text{i}\xi)e^{-\text{i}\phi }   & \text{i}\xi e^{\text{i}\phi}  \\
	-\text{i}\xi e^{-\text{i}\phi}  & (1-\text{i}\xi) e^{\text{i}\phi}  \\
\end{array}
\right)
\end{equation}	
and
\begin{equation}
	\xi=\frac{\Gamma _{\mathrm{eff}}}{2 \left(\Delta-\Delta _{\mathrm{L}}\right)}.
	\label{xi}
\end{equation}	 
The corresponding transmission and reflection amplitudes are of the form
\begin{subequations}
	\begin{equation}
t=\frac{e^{-\text{i}N\phi}}{\big(\tilde{\mathbf{T}}^N\big)_{11}},
		\label{finalt}
	\end{equation}	
	\begin{equation}
r=\frac{\big(\tilde{\mathbf{T}}^N \big)_{21}}{\big(\tilde{\mathbf{T}}^N\big) _{11}}.
		\label{finalr}
	\end{equation}
\end{subequations}

By using Abeles's theorem \cite{Abeles-AnnPhys1950}, we can derive the transmittance
and reflectance as follows  (see Appendix \ref{Derivation2} for details)
\begin{subequations}
	\begin{equation}
	T=|t|^2=\frac{1}{1+\xi ^2U^2_{N-1}\left( y \right)},
	\label{abst}
	\end{equation}	
	\begin{equation}
	R=|r|^2=\frac{\xi ^2U^2_{N-1}(y)}{1+\xi ^2U^2_{N-1}\left( y \right)}.		
	\label{absr}
	\end{equation}
\end{subequations}
Here $U_n(y)$ represents the Chebyshev polynomials of the second kind. Note that the transmittance and the reflectance are
constrained by the relation $T+R=1$ because of the conservation of photon number. Thus, in the following part, we focus on the
reflectance $R$ only.
\begin{figure*}[t]
	\centering
	\includegraphics[width=\textwidth]{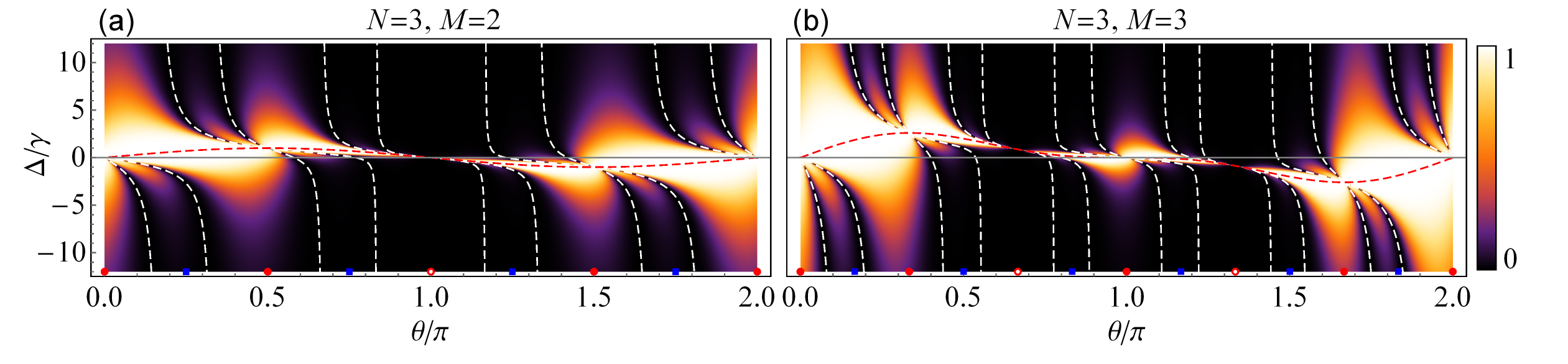}
		\caption{Reflectance $R$ for an array of separate giant atoms with maximum symmetry as functions of detuning $\Delta$
		and phase $\theta$, with (a) $N=3, M=2$, and (b) $N=3, M=3$. The red (white) dashed lines are used to mark the locations of the total 
		(zero) reflection. Some special phase delays are indicated by the red circles (decoupling), the red disks (superradiance), and 
		the blue squares (symmetrical broad-spectrum), respectively.
		}
		\label{SpectrumSeparate2D} 
\end{figure*}
\subsection{\label{SpectraSG}Spectra for the case of maximum symmetry}
Based on the results of the previous subsection, we further consider the case of maximum symmetry for an array of identical giant atoms in a separate configuration, as shown in Fig.~\ref{SeperateSketch}(c). In this case, all the bare decay rates are equal with $\gamma_{im}=\gamma$. In addition, the coupling points are equally spaced with distance $d$ [i.e., the lattice constant is $Md$, see Fig.~\ref{SeperateSketch}(c)], and the corresponding phase delay is $\theta=\omega_\mathrm{a} d/v_\mathrm{g}$. The effective decay rate and the Lamb shift of a single giant atom can be written as
\begin{subequations}
	\begin{equation}
\Gamma_{\mathrm{eff}}=
\gamma \frac{1-\cos M\theta}{1-\cos \theta},
		\label{Gammaeff1}
	\end{equation}
	\begin{equation}
\Delta_{\mathrm L}=\frac{\gamma}{2}\frac{M\sin\theta-\sin M\theta }{1-\cos \theta}
		\label{Deltaeff1},
	\end{equation}
\end{subequations}
and $y$ takes the form
\begin{equation}
	y=\cos M\theta+\xi \sin M \theta.
	\label{yMtheta}
\end{equation}
In Figs.~\ref{SpectrumSeparate2D}(a) and \ref{SpectrumSeparate2D}(b), we plot the reflectance as functions of the detuning $\Delta$ and the phase delay $\theta$ for $N=3$ and different $M$.  The range of $\theta$ is chosen as $\theta\in[0,2\pi]$ because the spectra change periodically with $\theta$.
From Eqs.~\eqref{abst} and \eqref{absr}, we can obtain that the reflection peaks with $R = 1$ appears at $\Delta=\Delta_{\mathrm{L}}$ [marked by the red dashed lines in Figs.~\ref{SpectrumSeparate2D}(a) and \ref{SpectrumSeparate2D}(b)]. And the reflection minima with $R=0$ can be fixed
by the relation $U_{N-1}[y(\Delta,\theta)]=0$ [marked by the white dashed lines in Figs.~\ref{SpectrumSeparate2D}(a) and \ref{SpectrumSeparate2D}(b), see more details in Sec.~\ref{ReflectionMinima}]. In addition, for a phase factor $\theta\in[0,\pi]$, we have relation $R(\Delta,\theta)=R(-\Delta,2\pi-\theta)$. The detailed characteristics of the spectra for different $\theta$ are summarized below.
\subsubsection{\label{Lorentz} $\theta=\frac{n\pi}{M}$: decoupling or superradiant points}
\begin{figure*}[t]
	\centering
	\includegraphics[width=0.75\textheight]{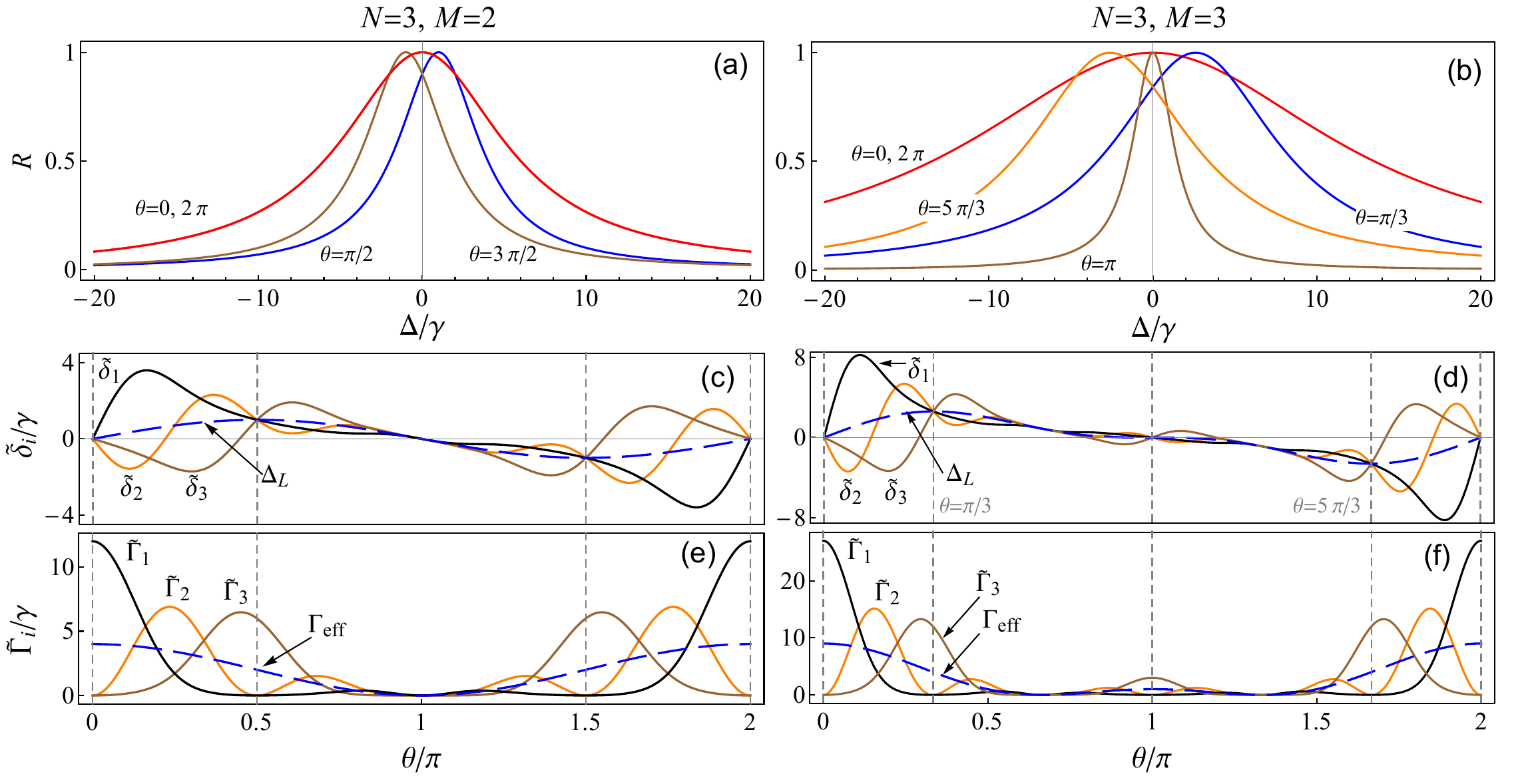}
	\caption{(a) and (b): Plots of $R$ vs $\Delta$ for an array of separate giant atoms with 
	maximum symmetry. The phase delays $\theta$ are chosen to satisfy the superradiant condition. (c) and (d): 
	The detuning $\tilde\delta_{i}$ between the $i$th collective mode and each atom (solid lines) and the Lamb shift 
	$\Delta_{\mathrm{L}}$ of each atom (blue dashed lines) as functions of $\theta$.
	(e) and (f): The decay rate $\tilde{\Gamma}_i$ of the $i$th mode (solid lines) and the effective decay 
	$\Gamma_{\mathrm{eff}}$ of each atom (blue dashed lines) as functions 
	of $\theta$. In panels (c)-(f), the dashed grid lines are used to mark the phase delays at which superradiance appears. 
	Left column: $N=3, M=2$. Right column: $N=3, M=3$.}
	\label{superradiance}
\end{figure*}
\begin{figure*}[t]
	\centering
	\includegraphics[width=0.75\textheight]{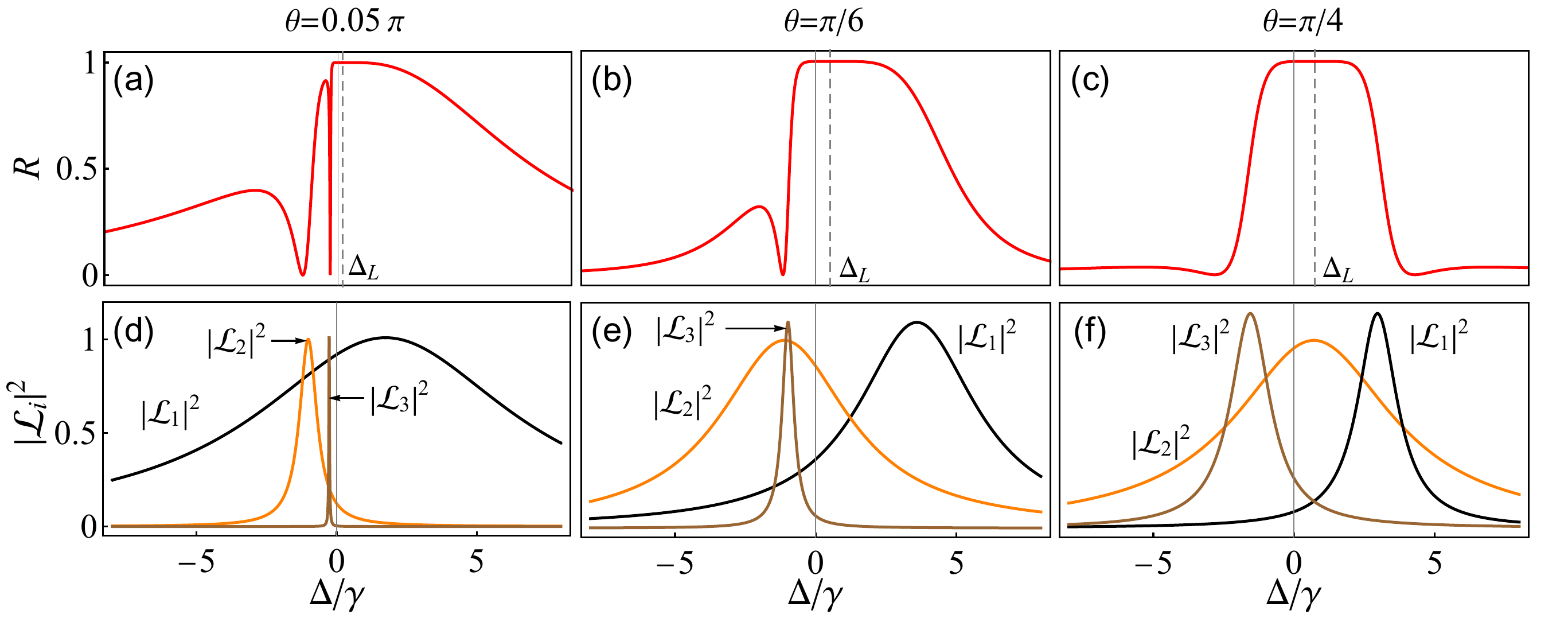}
	\caption{Panels (a)-(c) show reflection spectra with multiple reflection minima. In each panel, the Lamb shift of a single atom is indicated by 
	the dashed grid line where the reflection reaches its maximum. Corresponding to panels (a)-(c), panels (d)-(f) show the 
	decomposed components contributed by the collective excitations. In all the panels, we set $N=3, M=2$.}
	\label{FanoMinimum}
\end{figure*}
When $\theta={n\pi}/{M}$ ($n\in\mathbb{N}$), depending on the value of $n$, the atomic array decouples from the waveguide,
or exhibits a Lorenzian spectrum of width $N\Gamma_{\mathrm{eff}}$, which is a superradiant feature. 
   
\begin{itemize}
\item  If $n=2m$ and $\mathrm{mod}[m,M]\ne0$ ($m\in\mathbb{N}$), each atom is decoupled from the waveguide with $\Gamma_{\mathrm{eff}}=0$ [see Eq.~\eqref{Gammaeff1}], and hence $R\equiv 0$ for all values of $\Delta$ [see Eqs.~\eqref{xi} and
 \eqref{absr}]. Clearly, there are $M-1$ such decoupling points (with $n=2,4,\cdots,2M-2$) in the range of $\theta\in[0,2\pi]$, as shown
by the red circles in Figs.~\ref{SpectrumSeparate2D}(a) and \ref{SpectrumSeparate2D}(b). 

\item If $n=2m+1$, or $n=2m$ and $\mathrm{mod}[m,M]=0$
 ($m\in \mathbb{N}$), according to Eqs.~\eqref{Gammaeff1} and \eqref{Deltaeff1}, the effective decay rate and the Lamb shift become
 \begin{equation}
~~~~~~~~\Delta^{\mathrm{sup}} _{\mathrm{L}}=\left\{ 
\begin{array}{l}
	\frac{1}{2}M\gamma\cot\left(\frac{n}{2M}\pi\right),~~n=2m+1,
	\\
	\\
	0,~~n=2m~\mathrm{and}~\mathrm{mod}[m,M]=0,
\end{array} \right.
\label{SupDeltaL}
\end{equation}	
and
\begin{equation}
~~~~~~~~\Gamma^{\mathrm{sup}} _{\mathrm{eff}}=\left\{ 
\begin{array}{l}
	{\gamma }{\csc^2  \left(\frac{n}{2M}\pi\right)},~~n=2m+1,
	\\
	\\
	M^2\gamma,~~n=2m~\mathrm{and}~\mathrm{mod}[m,M]=0.
\end{array} \right.
\label{SupWidth}
\end{equation}	 
In addition, according to Eq.~\eqref{yMtheta}, we have $y=\pm 1$. Thus, using the identity \eqref{specialvalue1}, the reflection amplitude \eqref{absr} can be further simplified as 
\begin{equation}
	R=\frac{\frac{1}{4}\left({N \Gamma^{\mathrm{sup}} _{\mathrm{eff}}}\right){}^2}{\left(\Delta-\Delta^{\mathrm{sup}} _{\mathrm{L}}\right)^2+\frac{1}{4}\left({N \Gamma^{\mathrm{sup}} _{\mathrm{eff}}}\right){}^2}.
	\label{super}
\end{equation}
In this case, the reflection spectrum is a standard Lorentzian line shape centered at 
$\Delta=\Delta^{\mathrm{sup}} _{\mathrm{L}}$ and the line width $N\Gamma^{\mathrm{sup}} _{\mathrm{eff}}$
is proportional to the number of giant atoms, which is a typical superradiant phenomenon. 
In the range of $\theta \in [0,2 \pi]$, there are $M+2$ such points (with $n=0,1,3,\cdots,2M-3,2M-1,2M$), as shown by the red disks in Figs.~\ref{SpectrumSeparate2D}(a) and \ref{SpectrumSeparate2D}(b). The corresponding cross sections at these phases, which exhibit Lorentzian-type spectra, are shown by the curves in Figs.~\ref{superradiance}(a) and \ref{superradiance}(b).
\end{itemize}

Now we compare these results with the case of an array of equally spaced small atoms \cite{Mukhopadhyay-PRA2019}. (i) For small atoms, the superradiance appears when the phase delay between neighboring atoms is $n\pi$. While for equally spaced separate giant atoms, as shown above, the condition satisfied by the phase delay between neighboring atoms $M\theta=n\pi$ is necessary but not sufficient for the appearance of superradiance, because the atom decouples from the waveguide (with $\Gamma_{\mathrm{eff}}=0$) when $n=2m$ and $\mathrm{mod}[m,M]\ne0$.
(ii) For small atoms, the line center is also the resonance point of each single atom. While for separate giant atoms, the line center appears at $\Delta=\Delta^{\mathrm{sup}}_{\mathrm L}$ [see Eq.~\eqref{SupDeltaL}] because the frequency of each atom is shifted by $\Delta^{\mathrm{sup}}_{\mathrm L}$ due to the exchange of virtual photons between different connection points of the same atom. 
(iii) For small atoms, the width of the superradiant spectrum is always $N\gamma$. While for separate giant atoms, the width 
becomes $N\Gamma^{\mathrm{sup}}_{\mathrm{eff}}$, which depends on the phase delay, as shown in Eq.~\eqref{SupWidth}.

As shown by Eqs.~\eqref{tDecompose2} and \eqref{rDecompose2}, the scattering spectra are the result of interference between Lorentzian-type excitations of different collective modes of the atomic array. This point of view can provide an effective way
to explain the formation of these superradiant states. To this end, we plot the detuning $\tilde{\delta}_i$ between the $i$th collective mode and the atoms [see Figs.~\ref{superradiance}(c) and \ref{superradiance}(d)] and the decay rate $\tilde{\Gamma}_i$ of this mode [see Figs.~\ref{superradiance}(e) and \ref{superradiance}(f)] as functions of $\theta$.
One can see that at the superraidant points [indicated by the dashed grid lines in Figs.~\ref{superradiance}(c)-\ref{superradiance}(f)], the frequencies of the 
collective states are degenerate and equal to the shifted effective frequency of a single giant atom, with $\tilde{\delta}_i=\Delta^{\mathrm{sup}} _{\mathrm{L}}$ [given by Eq.~\eqref{SupDeltaL}]. Additionally, among these states, one is a superraidant state, with width $N\Gamma^{\mathrm{sup}} _{\mathrm{eff}}$ [given by Eq.~\eqref{SupWidth}], and 
the others are subradiant, with vanished decay rates.       
Thus only the superradiant state contributes to the scattering process, resulting in a Lorentzian line shape centered
at $\Delta=\Delta^{\mathrm{sup}} _{\mathrm{L}}$ with width $N\Gamma^{\mathrm{sup}} _{\mathrm{eff}}$, as shown in Figs.~\ref{superradiance}(a) and \ref{superradiance}(b). 

\subsubsection{\label{ReflectionMinima} $\theta\ne\frac{n\pi}{M}$: spectra with multiple reflection minima}
According to Eqs.~\eqref{yroot} and \eqref{yMtheta}, we can find $N-1$ reflection minima appearing at 
\begin{equation}
\Delta_s=\Delta_{\mathrm {L}}+\frac{\sin M\theta}{2\left(y_s-\cos M\theta\right)}\Gamma_{\mathrm{eff}}
\label{delta_s}
\end{equation}
when $\theta\neq\left(2m\pm\frac{s}{N}\right)\frac{\pi}{M}$ ($m\in\mathbb{N}$ and $s=1,2,\cdots,N-1$), as shown in Fig.~\ref{FanoMinimum}(a). $y_s$ are the roots of of Chebyshev polynomial $U_{N-1}(y)$ given in Eq.~\eqref{yroot}.
If the value of $\cos M\theta$ is exactly equal to one of the roots of $U_{N-1}(y)$, labeled as $y_{s'}$, i.e., $\theta=\left(2m\pm\frac{s'}{N}\right)\frac{\pi}{M}$ is satisfied, the corresponding $\Delta_{s'}$ does not exist because the second 
term of Eq.~\eqref{delta_s} is divergent. Thus the number of reflection minima for this case is $N-2$ [see Fig.~\ref{FanoMinimum}(b)]. In particular, when $\theta=(2m+1)\pi/(2M)$, one can obtain a broad spectrum structure that is symmetric at the Lamb shift [see Fig.~\ref{FanoMinimum}(c)]. And there are $N-1$ ($N-2$) reflection minima symmetrically distributed on both sides of the main peak for $N\in\mathbb{O}^{+}$ ($N\in\mathbb{E}^{+}$). 
In the range of $\theta\in [0,2 \pi]$, there are $2M$ points (with $m=0,1,2,\cdots,2M-3,2M-1$) that exhibit this kind of spectrum structure, as indicated by the blue squares in Figs.~\ref{SpectrumSeparate2D}(a) and \ref{SpectrumSeparate2D}(b). Note that in the large $N$ limit, this type of spectral structure becomes a photonic band gap, which will be discussed in detail in Sec.~\ref{PhotonicBG}.

To better understand the appearance of these reflection minima, based on Eq.~\eqref{rDecompose2} we decompose the reflection amplitude into the sum of several 
Lorentzian-type amplitudes (defined as $\mathcal{L}_i\equiv{\tilde\eta_i}/\big({\Delta-\tilde{\delta}_{i}+\mathrm{i}{\tilde{\Gamma}_{i}}/{2}}\big), i=1,2,\cdots,N$) corresponding to the excitations of the collective modes [see Figs.~\ref{FanoMinimum}(d)-\ref{FanoMinimum}(f)].  
It can be seen that the positions of the reflection minima are determined by the peaks of the narrow excitation amplitudes, indicating that the minima result from the destructive interference between the broad and narrow collective excitation spectra. 
If the linewidth difference between the broad and narrow spectral lines is large enough, the reflection spectra around the minima exhibit Fano line shapes.
\subsubsection{\label{PhotonicBG} Photonic band gap}
\begin{figure}[t]
	\centering
	\includegraphics[width=0.5\textwidth]{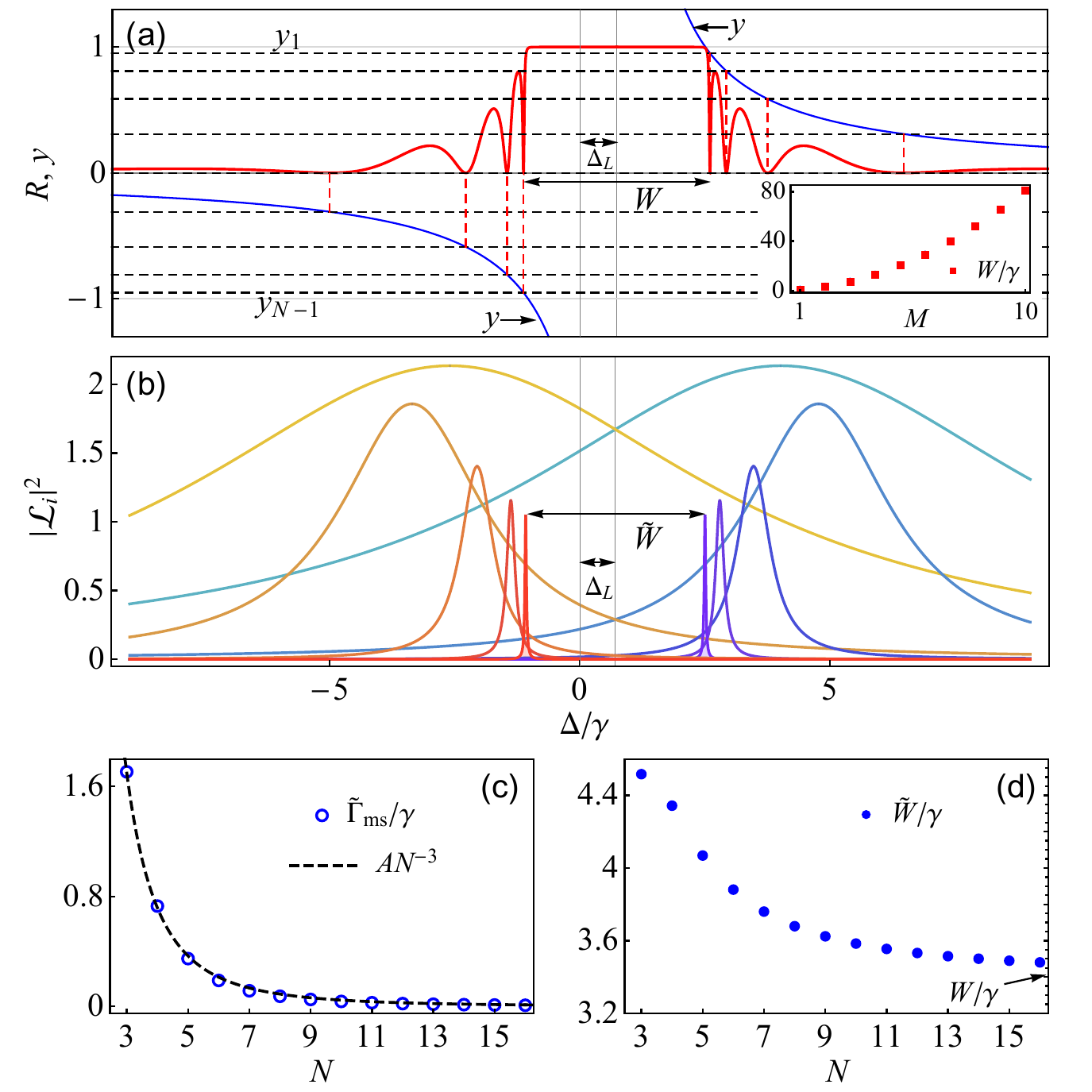}
	\caption{(a) The photonic band gap centered at $\Delta=\Delta_{\mathrm{L}}$ with width $W$. The parameters are 
	set as $N=10$, $M=2$, and $\theta=\pi/4$. The curve $y$ is plotted according to Eq.~\eqref{yMtheta}. The horizontal 
	coordinates $\Delta_s$ corresponding to $y=y_s$ ($s=1,2,\cdots,N-1$; and $s\neq N/2$ when $N\in\mathbb{E}^{+}$) are 
	the positions of zero reflection. When $N\gg 1$, the band-gap width can be defined as
	$W\simeq|\Delta_{1}-\Delta_{N-1}|$. The inset is the width $W$ for $\theta=\pi/(2M), 2\pi-\pi/(2M)$ (where $W$ 
	reaches the maximum for fixed $M$) as a function of $M$. (b) The curves show the 
	decomposed components contributed by the collective excitations corresponding to the spectra shown in panel (a). The 
	two narrowest peaks with distance $\tilde W$ and width $\tilde\Gamma_{\mathrm{ms}}$ are located at the innermost, 
	corresponding to the two most subradiant states. (c) The circles show the width $\tilde\Gamma_{\mathrm{ms}}$ as a 
	function of $N$ for fixed $M=2$. The solid line $AN^{-3}$ is the fitted curve with $A=45.94$. (d) The disks show the distance 
	$\tilde W$ as a function of $N$ for fixed $M=2$. It converges to $W$ for large $N$.}
	\label{bandgap}
\end{figure}
Interestingly, when the number of atoms $N$ is large, the broad spectrum structure appearing at $\theta=(2m+1)\pi/(2M)$ become a photonic band gap centered around the Lamb shift $\Delta_{\mathrm{L}}$, as shown in Fig.~\ref{bandgap}(a).  At the edges of the band gap, the reflectance drops sharply from 1 to 0. Thus, the distance between the two innermost reflection minima can be used to estimate the width of the band gap. In fact,
in the limit $N\gg1$, we have $y_1=-y_{N-1}\simeq 1$ [see Eq.~\eqref{yroot} and Fig.~\ref{bandgap}(a)], thus by using Eq.~\eqref{delta_s}, 
we can obtain the following approximate expression for the width of the band gap
\begin{equation}
W\simeq|\Delta_1-\Delta_{N-1}|\simeq\Gamma_{\mathrm{eff}}=\frac{\gamma }{1-\cos\left(\frac{2m+1}{2 M}\pi\right)}.
\label{Deltaw2}
\end{equation}
Clearly, in the range of $\theta\in[0,2\pi]$, for fixed $M$, $W$ reaches the maximum value when $m=0,2M-1$ [i.e., $\theta=\pi/(2M),2\pi-\pi/(2M)$].
The width $W$ for these phase delays is plotted as a function of $M$ in the inset of Fig.~\ref{bandgap}(a). When $M \gg 1$,  the maximum width can be further approximated as  
\begin{equation}
W\simeq\frac{8M^2}{\pi^2}\gamma,
\label{Deltaw2}
\end{equation}
indicating that $W$ increases as $M^2$ [see the inset in Fig.~\ref{bandgap}(a)]. Note that a similar band-gap structure of width $\gamma$ can also be created by an array of equally spaced small atoms with a phase delay of $(m+1/2)\pi$ \cite{Fang-PRA2015,He-OL2021}. But the giant-atom systems provide an effective way to significantly increase the width of the band gap by increasing the number of coupling points of each atom. In addition, the center of the photonic band gap is also tunable by changing the Lamb shift, which is impossible for an array of small atoms. It should also be emphasized that both the band-gap structure discussed here and the similar phenomenon in small atoms \cite{Fang-PRA2015,He-OL2021} appear in the Markovian regime. 
While Ref.~\cite{Greenberg-PRA2021} show that in the non-Markovian regime, the phase-accumulated effects for detuned photons can also produce a flat band gap structure in the transmission spectrum if the distance between neighboring qubits is equal to a half wavelength.  
	
The formation of this type of spectra can be understood as interference effects between different scattering channels corresponding to the excitations of the collective modes.
To this end, according to Eq.~\eqref{rDecompose2}, we decompose the spectra into the superposition of Lorentzian lines.  As shown in Fig.~\ref{bandgap}(b), these Lorentzian-type amplitudes are distributed symmetrically around the Lamb shift. The area of nearly total reflection is mainly contributed from the excitation of the most superradiant states. On the other hand, we find that the two narrowest excitation amplitudes
corresponding to the most subradiant states are at the innermost [see Fig.~\ref{bandgap}(b)], and the corresponding width $\tilde\Gamma_{\mathrm{ms}}$ decreases as $1/N^3$ [see Fig.~\ref{bandgap}(c)], resulting in very narrow peaks for large $N$. The destructive interferences between them and the broad-width excitations lead to the innermost reflection minima and the steep band-gap walls in the spectrum. Thus, when $N\gg 1$, the distance $\tilde W$ between the centers of the two innermost amplitudes converges to the width $W$ of the band gap, as shown in Fig.~\ref{bandgap}(d). 

\section{\label{BraidedDFI}Spectra for an array of braided giant atoms with decoherence-free interactions}
\begin{figure}[t]
	\centering
	\includegraphics[width=0.5\textwidth]{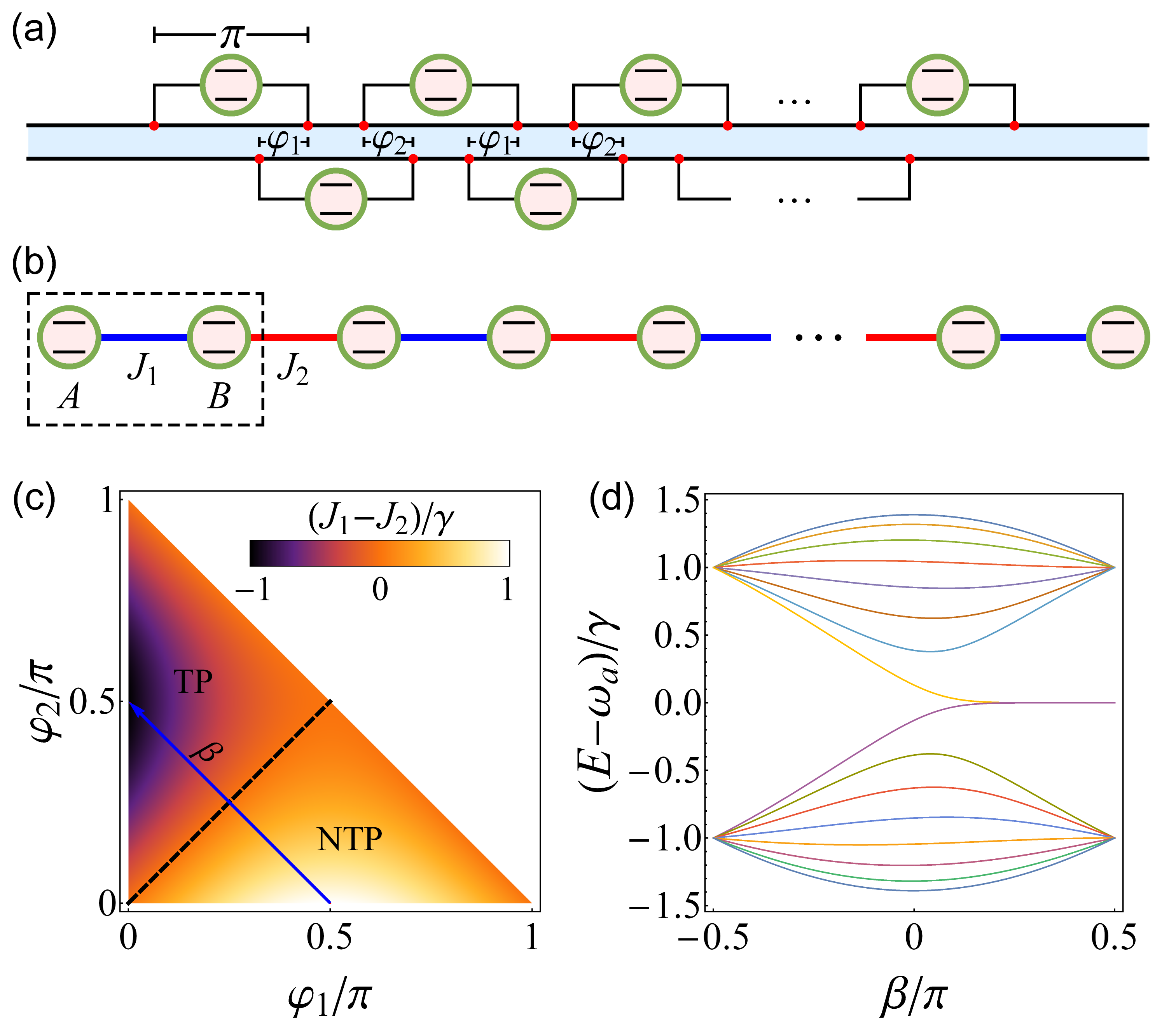}
	\caption{(a) Sketch of a setup with giant atoms realizing an SSH chain with decoherence-free nearest-neighbor couplings. (b) The 
	effective system. The dashed box shows the unit cell. (c) $J_1-J_2$ is plotted as a function of $\varphi_1$ and $\varphi_2$. There 
	are topological phase (TP) and non-topological phase (NTP) for different values of $\varphi_1$ and $\varphi_1$. The dashed line 
	($\varphi_1=\varphi_2$) indicates the topological phase transition where $J_1=J_2$ is satisfied. (d) Under the condition 
	$\varphi_1+\varphi_2=\pi/2$, the energy levels are plotted as functions of $\beta=\varphi_1-\varphi_2$ [indicated by the blue solid 
	line with arrow in (c)] for the setup shown in (a). The number of atoms is set as $N=16$.}
	\label{SSH}
\end{figure}
For an array of $N$ braided or nested giant atoms, the transfer-matrix approach is no longer available due to the feedback loops that exist in these configurations. Thus the general formulas Eqs.~\eqref{tGeneral} and \eqref{rGeneral} should be used to calculate the scattering coefficients. In general, when the atomic number is large, the explicit analytical expressions for the scattering coefficients are not simple, even tedious, except for the case $N=2$ \cite{Feng-PRA2021}. 
For the sake of compactness, in this section, we focus on the spectra that can reveal non-trivial many-body states  caused by decoherence-free interactions in an array of braided giant atoms. The spectra for an array of braided giant atoms in the other parameter regime, as well as the spectra for an array of nested giant atoms, are provided in the Appendixes \ref{SpectraBraided} and \ref{SpectraNested}.
\subsection{\label{SSHscheme} A scheme for constructing an SSH-type atomic chain using decoherence-free interactions} 
Here, we consider a 1D chain containing $N$ identical giant atoms, each with a transition frequency $\omega_{\mathrm{a}}$ and two connection points. Each pair of neighboring atoms is in a braided configuration, and all the bare decay rates are equal with $\gamma_{im}=\gamma$. For this configuration, the decoherence-free interactions between each pair of nearest neighboring atoms are easily designed \cite{Kockum-PRL2018}. Here, as an example, we construct a topological spin chain described by the SSH model \cite{Su-PRL1979}, which can be used to understand some of the fundamental ideas of topological physics \cite{Qi-RMP2011,Haldane-PRL2008,Ozawa-RMP2019,Meier-NatCommun2016,Lohse-NatPhys2016,Nakajima-NatPhys2016}.  To this end, we set the phase delay as $\theta_{i2}-\theta_{i1}=\pi$, and $\theta_{i2}-\theta_{i+1,1}=\varphi_1$ for $i\in\mathbb{O}^{+}$ and $\theta_{i2}-\theta_{i+1,1}=\varphi_2$ for $i\in\mathbb{E}^{+}$, as shown in Fig.~\ref{SSH}(a). 
Thus, from Eqs.~\eqref{Deltaeff}-\eqref{CollDecay}, we find that with such a design, the Lamb shifts, the individual decays, and the collective decays vanish, with $\Delta_{\mathrm{L},i}=\Gamma_{\mathrm{L},i}=\Gamma_{\mathrm{coll},ij}=0$. 
And for exchange interactions, only the nearest-neighbor ones are nonzero, with the so-called decoherence-free interactions $g_{i,i+1}=\gamma\sin\varphi_1\equiv J_1$ for $i\in\mathbb{O}^{+}$ and $g_{i,i+1}=\gamma\sin\varphi_2\equiv J_2$ for $i\in\mathbb{E}^{+}$, respectively.  
All other pairs of atoms do not interact with each other, so the tight-binding approximation is valid. Thus according to Eqs.~\eqref{Heff}, \eqref{EffHMarkovii}, and \eqref{EffHMarkovij}, we obtain the following effective Hamiltonian 
\begin{eqnarray}
\hat H_{\mathrm{SSH}}=&&\omega_{\mathrm{a}}\sum_{i=1}^{N} \hat\sigma^{+}_{i}\hat\sigma^{-}_{i}+J_{1}\sum_{i=\mathrm{odd}}\left(\hat\sigma^{+}_{i}\hat\sigma^{-}_{i+1}+\mathrm{H.c.}\right)
\nonumber
\\
&&+J_{2}\sum_{i=\mathrm{even}}\left(\hat\sigma^{+}_{i}\hat\sigma^{-}_{i+1}+\mathrm{H.c.}\right),
\label{H1DTop}
\end{eqnarray}
i.e., the setup is effectively a 1D chain of atoms described by the SSH model [see Fig.~\ref{SSH}(b), where the dashed box shows the unit cell].
The strengths of nearest-neighbor coupling are $J_1$ (for hopping within the unit cell) and $J_2$
(for hopping connecting neighboring unit cells), respectively.

By assuming periodic boundary conditions, one can obtain the corresponding bands 
\begin{equation}
E(K)=\omega_{\mathrm{a}}\pm\sqrt{{J_1^2}+{J_2^2}+2J_1J_2\cos K},
\end{equation}
where $K\in [-\pi,\pi]$ is the wave vector. The spectrum is gapped and forms two symmetric bands centered at the reference frequency $\omega_{\mathrm{a}}$, with the spectrum width and band gap 
\begin{subequations}
\begin{equation}
\Delta\omega=2(J_1+J_2)=4\gamma\sin\frac{\varphi_1+\varphi_2}{2}\cos\frac{\varphi_1-\varphi_2}{2},
\end{equation}
\begin{equation}  
\delta\omega=2\left|J_1-J_2\right|=4\gamma\cos\frac{\varphi_1+\varphi_2}{2}\left|\sin\frac{\varphi_1-\varphi_2}{2}\right|, 
\end{equation}
\end{subequations}
respectively. 

For finite systems and an even number of atoms,
the values of $J_{1,2}$ determine whether the system is in a topological phase (with  $J_1<J_2$, or $\varphi_1<\varphi_2$) or a non-topological
phase (with $J_1>J_2$, or $\varphi_1>\varphi_2$), as shown in Fig.~\ref{SSH}(c). Here we assume $\varphi_1+\varphi_2=\pi/2$ and let $\varphi_2-\varphi_1=\beta$ (i.e., $\varphi_{1,2}=\pi/4\mp\beta/2$, $\beta\in[-\pi/2,\pi/2]$), as indicated by the blue solid line with arrow in Fig.~\ref{SSH}(c). To show the influence of the
parameter $\beta$ on the topology of the system, we plot the energy levels of the atomic chain as functions of $\beta$ in Fig.~\ref{SSH}(d). 
The bandwith and the gap for this case are $\Delta\omega=2\sqrt{2}\gamma\cos({\beta}/{2})$ and $\delta\omega=2\sqrt{2}\gamma\left|\sin({\beta}/{2})\right|$, respectively.
We can
see that if $\beta<0$, no edge states appear in the gap. Therefore, the system is in a topologically trivial phase.
In the case of $\beta=0$, the gap closes, recovering the normal 1D tight-binding model. If
$\beta>0$, there are a pair of almost-zero-energy edge states $|\Psi_{\pm}\rangle$ in the spectrum gap, indicating that
the system is in a topologically non-trivial phase. If the energy-level splitting is tiny (This is valid when $\beta$ is not too small), 
$|\Psi_{\pm}\rangle$ take the form of the hybridization of the left and the right edge states $|\Psi_{\pm}\rangle=(|\Psi_{\mathrm{L}}\rangle\pm|\Psi_{\mathrm{R}}\rangle)/{\sqrt{2}}$, with
\begin{subequations}
\begin{equation}
\left|\Psi_{\mathrm{L}}\right\rangle=\frac{1}{\sqrt{\mathcal{N}_{\mathrm{L}}}}\sum_{i=\mathrm{odd}}\left(-\mu\right)^{\frac{i-1}{2}}\left|i\right\rangle,
\label{LState}
\end{equation} 
\begin{equation}
\left|\Psi_{\mathrm{R}}\right\rangle=\frac{1}{\sqrt{\mathcal{N}_{\mathrm{R}}}}\sum_{i=\mathrm{even}}\left(-\mu\right)^{\frac{N-i}{2}}\left|i\right\rangle.
\label{RState}
\end{equation} 
\end{subequations}
Here $\mu=J_1/J_2\in(0,1)$, $\left|i\right\rangle=\hat\sigma_{i}^{+}\left|\mathrm{G}\right\rangle$ is the excited state of the $i$th atom, and $\mathcal{N}_{\mathrm{R,L}}$ is the normalization constant. 
$|\Psi_{\mathrm{L}}\rangle$ and $|\Psi_{\mathrm{R}}\rangle$ hybridize under $\hat{H}_{\mathrm{SSH}}$ to an exponentially small amount, resulting in an effective interaction 
\begin{equation}
\mathcal{J}\simeq J_2\left(\mu^2-1\right)\left(-\mu\right)^{\frac{N}{2}}
\label{JRL}
\end{equation} 
between them. Thus the splitting between $|\Psi_{+}\rangle$ and $|\Psi_{-}\rangle$ is approximately $2|\mathcal{J}|$. 
\begin{figure*}[t]
	\centering
	\includegraphics[width=\textwidth]{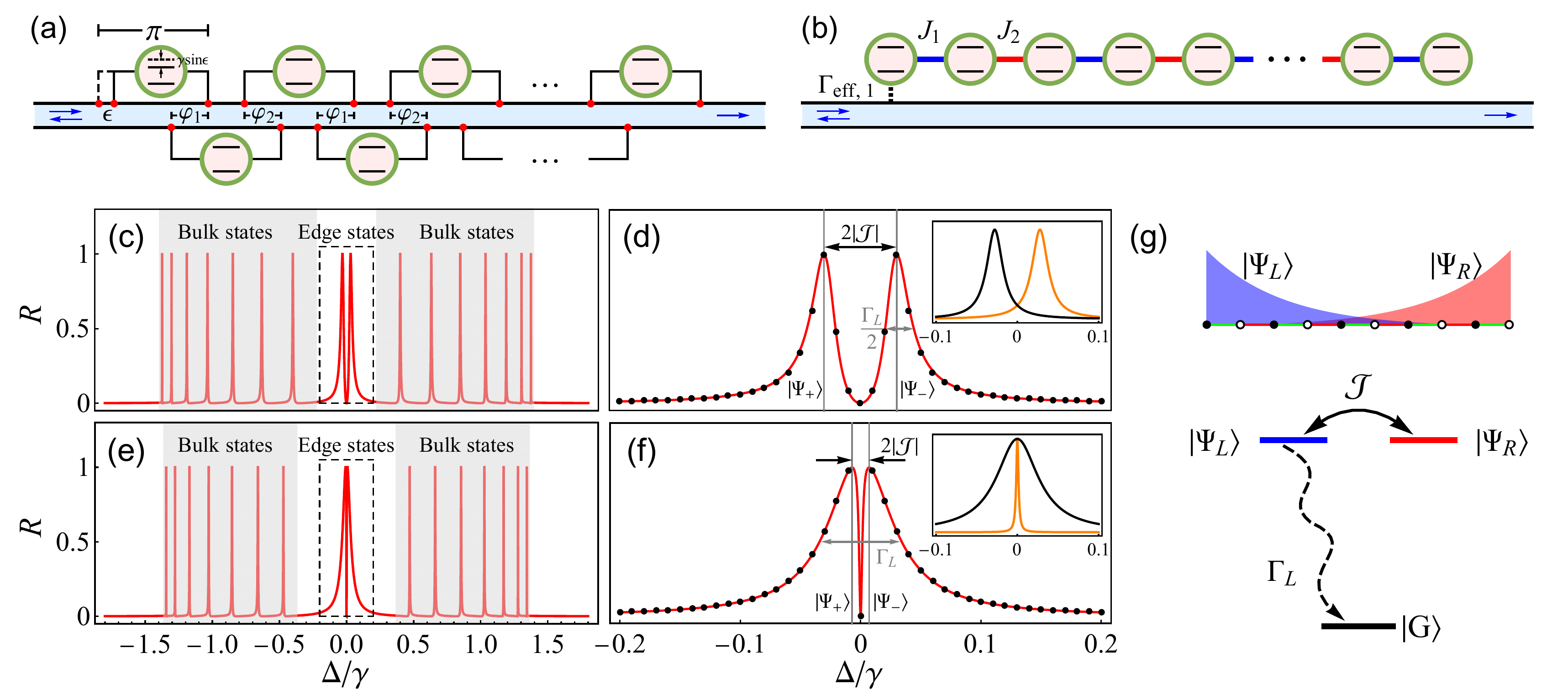}
	\caption{(a) Scheme for probing the decoherence-free interactions of an atomic array by properly designing the first atom. (b) The effective 
	system. (c)-(f) The reflection spectra for an array of $N=16$ braided atoms with decoherence-free interactions. The bands formed
	by the bulk states in the thermodynamic limit $N\to\infty$ are shown by the shaded regions in (c) and (e). 
	Panels (d) and (f) show the details of the spectral structures in the dashed boxes in panels (c) and (e), corresponding to the edge states. 
	The black dots in these panels are approximate results obtained from Eq.~\eqref{rtophap}.
	The insets in panels (d) and (f) show the corresponding decomposed components $|{\tilde\eta_{\pm}}/{(\Delta-Z_{\pm})}|^2$ contributed by 
	the collective excitations. The parameters 
	are $\varphi_1=0.2\pi$, $\varphi_2=0.3\pi$, and $\epsilon=0.1\pi$ in (c) and (d); $\varphi_1=\pi/6$, $\varphi_2=\pi/3$, 
	and $\epsilon=0.1\pi$ in (e) and (f). (g) Upper panel: The superposition of the left edge state $|\Psi_{\mathrm{L}}\rangle$ and the right edge 
	state $|\Psi_{\mathrm{R}}\rangle$ results in an effective coupling strength $\mathcal{J}$ between them. Lower panel: Effective energy 
	diagram formed by the edge states $|\Psi_{\mathrm{L}}\rangle$, $|\Psi_{\mathrm{R}}\rangle$, and the ground state $|\mathrm{G}\rangle$.}
	\label{SSHspectrum}
\end{figure*}
\subsection{\label{probeSSH} Probe the energy spectrum of an SSH-type atomic chain formed by decoherence-free interactions} 
The exact decoherence-free interactions described by Eq.~\eqref{H1DTop} cannot be probed by the photons because all the atoms are decoupled from 
the waveguide. To overcome this problem, we can slightly change the position of the leftmost coupling point, with $\theta_{11}\to\theta_{11}+\epsilon$. Namely, the phase delay between the two connection points of the first atom becomes $\pi-\epsilon$, as shown by Fig.~\ref{SSHspectrum}(a). This design can ensure that all the exchange interactions and the collective decays between the first and the other atoms remain unchanged [i.e., $g_{12}=J_1=\gamma\sin\varphi_1$, $g_{1j}=0$ ($j\ge 3$), and $\Gamma_{\mathrm{coll},1j}=0$  ($j\ge 2$)], 
and at the same time, the individual decay and the Lamb shift of the first atom obtain small values $\Gamma_{\mathrm{eff},1}=2\gamma(1-\cos\epsilon)$ and $\Delta_{\mathrm{L},1}=\gamma\sin\epsilon$, 
respectively [see Eqs.~\eqref{Deltaeff}-\eqref{CollDecay}]. As shown by Fig.~\ref{SSHspectrum}(a), we also let the transition frequency of the first atom be $\omega_{1}=\omega_{\mathrm{a}}-\gamma\sin\epsilon$ to cancel out the influence
of the Lamb shift $\Delta_{\mathrm{L},1}$ (i.e., all the effective atomic frequencies equal to $\omega_{\mathrm{a}}$). As a result, 
the effective Hamiltonian \eqref{H1DTop} of the system becomes
\begin{equation}
\hat H'_{\mathrm{eff}}=\hat H_{\mathrm{SSH}}-\frac{\mathrm{i}}{2}\Gamma_{\mathrm{eff},1}\hat\sigma^{+}_{1}\hat\sigma^{-}_{1},
\label{H1DTopWG}
\end{equation} 
which means that the system sketched in Fig.~\ref{SSHspectrum}(a) is equivalent to the one shown in Fig.~\ref{SSHspectrum}(b). Thus through the coupling between the waveguide and the first atom, the energy spectra of the topological atomic
chain can be probed. The corresponding scattering amplitudes can be calculated by using Eqs~\eqref{tGeneral} and \eqref{rGeneral}.

The reflection spectra of the system in the topologically non-trivial phase [with $\beta\in(0,\pi/2)$] are shown in Figs.~\ref{SSHspectrum}(c)-\ref{SSHspectrum}(f). 
It can be seen that a mapping relationship is established between the reflection peaks and the energy levels of the atomic array [see Figs.~\ref{SSHspectrum}(c) and \ref{SSHspectrum}(e)]. On one hand,
the peaks corresponding to the bulk states are very narrow and thus well resolved [see the spectra in the shaded regions in Figs.~\ref{SSHspectrum}(c) and \ref{SSHspectrum}(e)], since the effective decays of these states to the 
waveguide are small. On the other hand, the properties of the spectral structure around $\Delta=0$, corresponding to the hybridized edge states $|\Psi_{\pm}\rangle$, depend on $\beta$. When $\beta$ is small, the spectrum around $\Delta=0$ exhibits the Autler-Townes splitting (ATS) feature, as shown in Figs.~\ref{SSHspectrum}(c) and \ref{SSHspectrum}(d). While for large $\beta$, the spectrum has the character of EIT, as shown in Figs.~\ref{SSHspectrum}(e) and \ref{SSHspectrum}(f). 

To further understand the spectral structure in the topological band gap, we assume that $\beta$ is not too small so that the edge states are deep in the topological band gap and the splitting between them is tiny. Because of the existence of the topological band gap, the bulk states are almost not coupled to the edge states and can therefore be ignored. Therefore, we can only consider the subspace spanned by $|\Psi_{\mathrm{L,R}}\rangle$ and $|\mathrm{G}\rangle$, in which the effective non-Hermitian Hamiltonian \eqref{H1DTopWG} can be 
approximated as  
\begin{equation}
\hat H_{\mathrm{RL}}\simeq\omega_{\mathrm{a}}\sum_{i=\mathrm{L,R}}\hat{S}^{\dag}_{i}\hat{S}_{i}-\frac{\mathrm{i}}{2}\Gamma_{\mathrm{L}}\hat{S}_{\mathrm{L}}^{\dag}\hat{S}_{\mathrm{L}}
+\mathcal{J}\left(\hat{S}^{\dag}_{\mathrm{R}}\hat{S}_{\mathrm{L}}+\mathrm{H}.\mathrm{c}.\right),
\label{HRL}
\end{equation} 
where the lower operator is defined as $\hat{S}_{\mathrm{L,R}}=|\mathrm{G}\rangle\langle \Psi_{\mathrm{L,R}}|$.
The two edge states interact with strength $\mathcal{J}$, defined by Eq.~\eqref{JRL}. The left edge state $|\Psi_{\mathrm{L}}\rangle$ 
couples to the waveguide with an effective decay 
\begin{equation}
\Gamma_{\mathrm{L}}\simeq\left(1-\mu^2\right)\Gamma_{\mathrm{eff},1},
\label{GammaL}
\end{equation} 
resulting from the coupling between the leftmost atom and the waveguide. And the right edge state $|\Psi_{\mathrm{R}}\rangle$ is 
decoupled from the waveguide with effective decay $\Gamma_{\mathrm{R}}=0$. Thus the left and right
edge states $|\Psi_{\mathrm{L,R}}\rangle$ and the ground state $|\mathrm{G}\rangle$ form an effective three-level $\Lambda$-type system [see 
Fig.~\ref{SSHspectrum}(g)] that can exhibit ATS- or EIT-like spectra.

According to the effective Hamiltonian~\eqref{HRL} and Eqs.~\eqref{tDecompose2}-\eqref{rDecompose2}, the transmission and reflection amplitudes around the center of the band gap $\Delta=0$ can be approximated as
\begin{subequations}
	\begin{equation}
		t \simeq 1+\sum_{n=\pm}\frac{\eta_n}{\Delta-Z_{n}}=\frac{\Delta^2-\mathcal{J}^2}{\Delta\left(\Delta+\mathrm{i}\frac{\Gamma_{\mathrm{L}}}{2}\right)-\mathcal{J}^2},
		\label{ttophap}
	\end{equation}	
	\begin{equation}
		r \simeq \sum_{n=\pm}\frac{\tilde\eta_n}{\Delta-Z_{n}}=\frac{\mathrm{i}\frac{\Gamma_{\mathrm{L}}}{2}\Delta e^{\mathrm{i}\epsilon}}{\Delta\left(\Delta+\mathrm{i}\frac{\Gamma_{\mathrm{L}}}{2}\right)-\mathcal{J}^2},
		\label{rtophap}
	\end{equation}	
\end{subequations}
where 
\begin{equation} 
Z_{\pm}=\frac{1}{4}\left(-\mathrm{i}\Gamma_{\mathrm{L}}\pm\sqrt{16\mathcal{J}^2-\Gamma_{\mathrm{L}}^2}\right)
\end{equation}
is the eigenvalue of the non-Hermitian Hamiltonian Eq.~\eqref{HRL}. 
\begin{subequations}
	\begin{equation}
		\eta_{\pm}=\frac{\left(Z_{+}+Z_{-}\right)Z_{\pm}}{Z_{\pm}-Z_{\mp}},
		\label{tildeeta}
	\end{equation}	
	\begin{equation}
		\tilde\eta_{\pm}=-e^{\mathrm{i}\epsilon}\frac{\left(Z_{+}+Z_{-}\right)Z_{\pm}}{Z_{\pm}-Z_{\mp}}
		\label{eta}
	\end{equation}	
\end{subequations}
are weighting factors of the Lorentzian components. 
We can see from Figs.~\ref{SSHspectrum}(d) and \ref{SSHspectrum}(f) that the approximate results (the black dots) obtained from
Eq.~\eqref{rtophap} are in good agreement with the exact solutions (the red solid lines).
The ratio $4|\mathcal{J}|/\Gamma_{\mathrm{L}}$, which decreases monotonically in the region $\beta\in(0,\pi/2)$, can be used to determine whether the system is in the ATS or EIT region. 
When $4|\mathcal{J}|/\Gamma_{\mathrm{L}}>1$, the eigenvalues $Z_{\pm}$ have the same 
imaginary part and the opposite real parts. The system is thus in the ATS regime, where the spectral structure in the gap can be 
decomposed into two symmetrically distributed resonances of equal width [see the inset in Fig.~\ref{SSHspectrum}(d)]. In 
particular, when $4|\mathcal{J}|/\Gamma_{\mathrm{L}}\gg 1$, the splitting $2|\mathcal{J}|$ between the two peaks corresponding to the 
hybridized edge states $|\Psi_{\pm}\rangle=(|\Psi_{\mathrm{L}}\rangle\pm|\Psi_{\mathrm{R}}\rangle)/{\sqrt{2}}$ [note that $|\Psi_{\pm}\rangle$ corresponds to $Z_{\pm}$ ($Z_{\mp}$) for $\mathcal{J}>0$ ($\mathcal{J}<0$) and plays the role of 
dressed state] is much larger than the width $\Gamma_{\mathrm{L}}/2$ of the peaks. Thus the two edge states 
can be well detected by probe photons, as shown in Fig.~\ref{SSHspectrum}(d). On the other hand, in the EIT regime with 
$4|\mathcal{J}|/\Gamma_{\mathrm{L}}<1$, $Z_{\pm}$ are purely imaginary but with different moduli. 
Thus the spectral structure in the gap can be decomposed into a wide and a narrow resonances, both centered at $\Delta=0$ [see 
the inset in Fig.~\ref{SSHspectrum}(f)]. 
The Fano-type destructive interference between them produces a reflection spectrum with EIT-type transparency points at 
$\Delta=0$. In particular, when $4|\mathcal{J}|/\Gamma_{\mathrm{L}}\ll 1$, a narrow transparency dip of width about 
$4\mathcal{J}^2/\Gamma_{\mathrm{L}}$ appears in a Lorenzian resonance of width $\Gamma_{\mathrm{L}}$, as shown in 
Fig.~\ref{SSHspectrum}(f). 

\section{\label{conclusion}conclusion}
In summary, we study the single-photon scattering problem in multi-giant-atom wQED systems. 
It is shown that the scattering spectra are determined by the characteristic quantities relevant to the photon exchange and interference effects between coupling points, including the Lamb shifts and the effective decays of single atoms, and the exchange interactions and the collective decays between different atoms.  
The obtained analytical expressions provide a clear physical picture of the multi-channel scattering process via collective excitations.
For separate giant atoms, we find that the transfer-matrix method applicable for small atoms can be generalized to this case due to the feature of the cascade scattering process. Using the above theoretical tools, we systematically investigate the characteristics of the scattering spectra for multi-giant-atom systems in different configurations. It is shown that interactions between photons and the collective modes of the atomic array result in abundant interesting phenomena, such as superradiance, Fano-type interference, and photonic band gap. And these phenomena are strongly influenced by the non-dipole and interference effects resulting from the phase delays between different connection points.   
We also take the SSH-type model as an example to investigate the optical properties of topological states resulting from the decoherence-free 
interactions of a chain of braided atoms. It is shown that the scattered photons can be used to probe the 
non-trivial many-body states of an array of giant atoms. In addition, the quantum interferences between the excitation channels of the nearly degenerate edge modes can generate 
topologically protected EIT-type phenomena.
Our study can provide insight into the scattering physics when photons interact with multiple giant atoms, and may provide powerful tools for manipulating photon transport in future quantum networks.

\begin{acknowledgments}
We thank Q. Y. Cai and W. Nie for helpful discussions. This work was supported by the National Natural Science Foundation of China (NSFC) under Grants No. 61871333, and No. 12147208.
\end{acknowledgments}

\appendix
\begin{widetext}
\section{\label{Derivation1} Derivation of general expressions for scattering amplitudes}
Equations \eqref{EoM1}–\eqref{EoM2} yield the following the boundary conditions 
\begin{subequations}
	\begin{equation}
-\mathrm i v_{\mathrm{g}}\left[ \Phi _{\mathrm{R}}(x_{im+})-\Phi_{\mathrm{R}}(x_{im-})\right] +V_{im}f_{i}=0,
		\label{B1}
\end{equation}	
\begin{equation}
\mathrm i v_{\mathrm{g}}\left[\Phi_{\mathrm{L}}(x_{im+})-\Phi _{\mathrm{L}}(x_{im-})\right] +V_{im}f_{i}=0,
		\label{B2}
\end{equation}
\begin{equation}
\Delta _{i}f_{i}-\frac{1}{2}\sum_{s}\sum_{m=1}^{M_i}V_{im}\left[ \Phi_{s}(x_{im+})+ \Phi _{s}(x_{im-})\right] =0.
\label{B3}
\end{equation}
\end{subequations}
Here $s=\mathrm{L}, \mathrm{R}$. $\Delta_i=\omega-\omega_i$ is the detuning between the photons
and the $i$th atom. To obtain Eqs.~\eqref{B1}-\eqref{B3}, we have used the following relations:  
\begin{subequations}
\begin{equation}
\int_{x_{im,-}}^{x_{im,+}}\frac{\partial}{\partial x}\Phi_{s}(x)\mathrm{d}x=\Phi_{s}(x_{im,+})-\Phi_{s}(x_{im,-}).
\end{equation}
\begin{equation}
\int_{x_{im,-}}^{x_{im,+}}\sum_{i'=1}^{N}\sum_{m'=1}^{M_{i'}} V_{i' m'} \delta\left(x-x_{i'm'}\right) f_{i'}\mathrm{d}x
=\sum_{i'=1}^{N}\sum_{m'=1}^{M_{i'}}V_{i' m'} f_{i'}\delta_{ii'}\delta_{mm'}=V_{i m} f_{i}.
\end{equation}
\begin{equation}
\Phi_{s}(x_{im}) =\frac{1}{2}\left[\Phi_{s}(x_{im,+}) +\Phi_{s}(x_{im,-})\right].
\end{equation}
\end{subequations}

Substituting Eqs.~\eqref{PhiR} and \eqref{PhiL} into Eqs.~\eqref{B1}-\eqref{B3}, we arrive at 
\begin{subequations}
\begin{equation}
-\mathrm i v_{\mathrm{g}}(t_{p _{im}}-t_{p _{im}-1})e^{\mathrm ikx_{im}}+V_{im}f_{i}=0,
		\label{Eot}
\end{equation}	
\begin{equation}
\mathrm i v_{\mathrm{g}}(r_{p _{im}+1}-r_{p _{im}})e^{-\mathrm ikx_{im}}+V_{im}f_{i}=0.
\label{Eor}
\end{equation}
\begin{equation}
\Delta _{i}f_{i}-\frac{1}{2}\sum _{m=1}^{M_i}V_{im}\left[(t_{p _{im}}+t_{p _{im}-1})e^{\mathrm i kx_{im}}+(r_{p _{im}+1}+r_{p _{im}})e^{-\mathrm i kx_{im}}\right]=0.
	\label{Eof}
\end{equation}
\end{subequations}
Here we have assumed that the $m$th coupling point of the $i$th atom, located at $x_{im}$, is the $p_{im}$th one of all the coupling points (counting from the first coupling point at the far left). Thus we have the following relations: $x_{p_{im}}=x_{im}$, $V_{p_{im}}=V_{im}$, and $f_{p_{im}}=f_{i}$.

Starting from Eqs.~\eqref{Eot} and \eqref{Eor}, through iteration (note that $t_0$=1, $r_{N_\mathrm{c}+1}=0$), we obtain the following relations
\begin{subequations}
	\begin{equation}
	t_{p_{im}}=1-\mathrm{i}\frac{1}{v_{\mathrm{g}}}\sum _{p^{\prime}=1}^{p_{im}}V_{p^{\prime}}e^{-\mathrm i kx_{p^{\prime}}}
	f_{p^{\prime}},
	\label{tp}
	\end{equation}	
	\begin{equation}
	r_{p_{im}}=-\mathrm{i}\frac{1}{v_{\mathrm{g}}}\sum_{p^{\prime}=p_{im}}^{N_{\mathrm{c}}}V_{p^{\prime}}e^{\mathrm ikx_{p^{\prime}}}
	f_{p^{\prime}}.
	\label{rp}
	\end{equation}
\end{subequations}
Thus the transmission and reflection amplitudes read
\begin{subequations}
\begin{equation}
		t=t_{N_\mathrm{c}}=1-\mathrm i \frac{1}{v_{\mathrm{g}}}\sum _{p^{\prime}=1}^{N_\mathrm{c}}V_{p^{\prime}}e^{-\mathrm i kx_{p^{\prime}}}f_{p^{\prime}}=1-\mathrm i \frac{1}{v_{\mathrm{g}}}\sum_{i=1}^{N}\sum_{m=1}^{M_i}V_{im}e^{-\mathrm i kx_{im}}f_{i},
		\label{inout_t}
\end{equation}	
\begin{equation}
		r=r_1=-\mathrm i \frac{1}{v_{\mathrm{g}}}\sum _{p^{\prime}=1}^{N_\mathrm{c}}V_{p^{\prime}}e^{\mathrm ikx_{p^{\prime}}}f_{p^{\prime}}=-\mathrm i \frac{1}{v_{\mathrm{g}}}\sum_{i=1}^{N}\sum_{m=1}^{M_i}V_{im}e^{\mathrm i kx_{im}}f_{i}.
		\label{inout_r}
\end{equation}
\end{subequations}
Substituting the expressions \eqref{tp} and \eqref{rp} into Eq.~\eqref{Eof}, we arrive at
\begin{eqnarray}
\Delta_{i}f_{i}-\frac{1}{2}\sum_{m=1}^{M_i}V_{im}\left[\left(2-\mathrm i \frac{1}{v_{\mathrm{g}}}\sum _{p^{\prime}=1}^{p _{im}}V_{p^{\prime}}e^{-\mathrm i kx_{p^{\prime}}}f_{p^{\prime}}-\mathrm i \frac{1}{v_{\mathrm{g}}}\sum _{p^{\prime}=1}^{p _{im}-1}V_{p^{\prime}}e^{-\mathrm i kx_{p^{\prime}}}f_{p^{\prime}}\right) e^{\mathrm i kx_{im}}\right.
\nonumber
\\
\left.{+\left(-\mathrm i \frac{1}{v_{\mathrm{g}}}\sum _{p^{\prime}=p_{im}+1}^{N_{\mathrm c}}V_{p^{\prime}}e^{\mathrm{i}kx_{p^{\prime}}}
f_{p^{\prime}} -\mathrm i \frac{1}{v_{\mathrm{g}}}\sum _{p^{\prime}=p_{im}}^{N_{\mathrm c}}V_{p^{\prime}}e^{\mathrm{i}kx_{p^{\prime}}}
f_{p^{\prime}}\right)e^{-\mathrm{i}kx_{im}}}\right]=0,
\end{eqnarray}
which can be simplified as 
\begin{equation}
\Delta _{i}f_{i}+\frac{\mathrm{i}}{v_\mathrm{g}}\sum_{m=1}^{M_i}\left[\sum _{p^{\prime}=1}^{p _{im}-1}V_{im}V_{p^{\prime}}e^{\mathrm ik (x_{im}-x_{p^{\prime}})}f_{p^{\prime}} +\sum _{p^{\prime}=p _{im}}^{N_\mathrm{c}}V_{im}V_{p^{\prime}}e^{\mathrm ik (x_{p^{\prime}}-x_{im})}f_{p^{\prime}} \right]=\sum_{m=1}^{M_i}V_{im}e^{\mathrm i kx_{im}},
\end{equation}
namely,
\begin{equation}
\Delta _{i}f_{i}+\frac{\mathrm i}{v_{\mathrm{g}}}\sum_{m=1}^{M_i}\sum _{p^{\prime}=1}^{N_{\mathrm{c}}}V_{im}V_{p^{\prime}}e^{\mathrm i k|x_{im}-x_{p^{\prime}}|}f_{p^{\prime}}=\sum_{m=1}^{M_i}V_{im}e^{\mathrm i kx_{im}}.
\label{fE1}
\end{equation}
Note that 
\begin{equation}
\sum _{p^{\prime}=1}^{N_{\mathrm{c}}}V_{p^{\prime}}e^{\mathrm i k|x_{im}-x_{p^{\prime}}|}f_{p^{\prime}}=\sum_{j=1}^{N}\sum _{m'=1}^{M_j}V_{jm'}e^{\mathrm i k|x_{im}-x_{jm'}|}f_{j}, 
\end{equation}
thus Eq.~\eqref{fE1} can be further written as
\begin{equation}
\Delta _{i}f_{i}+ \frac{\mathrm i}{v_{\mathrm{g}}}\sum_{j=1}^{N}\sum_{m=1}^{M_i}\sum_{m'=1}^{M_j}V_{im}V _{jm'} e^{\mathrm i k|x_{im}-x_{jm'}|}f_{j}=\sum_{m=1}^{M_i}V_{im}e^{\mathrm i kx_{im}},
\label{fE2}
\end{equation}	
By defining decay rate into the waveguide continuum $\gamma_{im}= {2V_{im}^{2}}/{v_{\mathrm{g}}}$ (through coupling point $x_{im}$) and phase delay $\theta_{im}(\omega)=kx_{im}=\omega x_{im}/v_{\mathrm{g}}$, we obtain the following linear equations for the atomic excitation amplitudes $f_{i}$ 
\begin{equation}
\Delta _{i}f_{i}+ \frac{ \mathrm i}{2}\sum_{j=1}^{N}\sum_{m=1}^{M_i}\sum_{m'=1}^{M_j}\sqrt{\gamma _{im}\gamma _{jm'}}e^{\mathrm{i}\left| \theta _{im}(\omega)-\theta _{jm'}(\omega) \right|} f_{j}=\sum_{m=1}^{M_i}{\sqrt{\frac{v_{\mathrm{g}}\gamma _{im}}{2}}}e^{\mathrm{i}\theta_{im}(\omega)},
\end{equation}	
or in a more compact form
\begin{equation}
\left(\omega \mathbf{I}-\mathbf{H}\right)\mathbf{f}=\sqrt{v_g}\mathbf{V}.
\label{fE3}
\end{equation}	
The transmission and reflection amplitudes Eqs.~\eqref{tp} and \eqref{rp} can be further written as 
\begin{subequations}
\begin{equation}
t=1-\mathrm i  \frac{1}{\sqrt{v_{\mathrm{g}}}}\sum _{ i}\sum _{m}\sqrt{\frac{\gamma _{im}}{2}}e^{-\mathrm i \theta_{im}(\omega)}f_{i}\equiv 1-\frac{\mathrm{i}}{\sqrt{v_{\mathrm{g}}}}\mathbf{V}^{\dag}\mathbf{f},
\label{tvector}
\end{equation}
\begin{equation}
r=-\mathrm i \frac{1}{\sqrt{v_{\mathrm{g}}}}\sum_{ i}\sum_{m}\sqrt{\frac{\gamma_{im}}{2}}e^{\mathrm i \theta_{im}(\omega)}f_{i}\equiv -\frac{\mathrm{i}}{\sqrt{v_{\mathrm{g}}}}\mathbf{V}^{\top}\mathbf{f}.
\label{rvector}
\end{equation}
\end{subequations}
Here $\mathbf{f}$, $\mathbf{V}$, and $\mathbf{H}$ are defined in Eqs.~\eqref{fmatrix}, \eqref{Vmatrix} and \eqref{EffH} in the main text.
From Eqs.~\eqref{fE3}-\eqref{rvector}, one can easily obtain the formal solutions Eqs.~\eqref{fGeneral}-\eqref{rGeneral}.
\end{widetext}
\begin{figure}[t]
	\centering
	\includegraphics[width=0.5\textwidth]{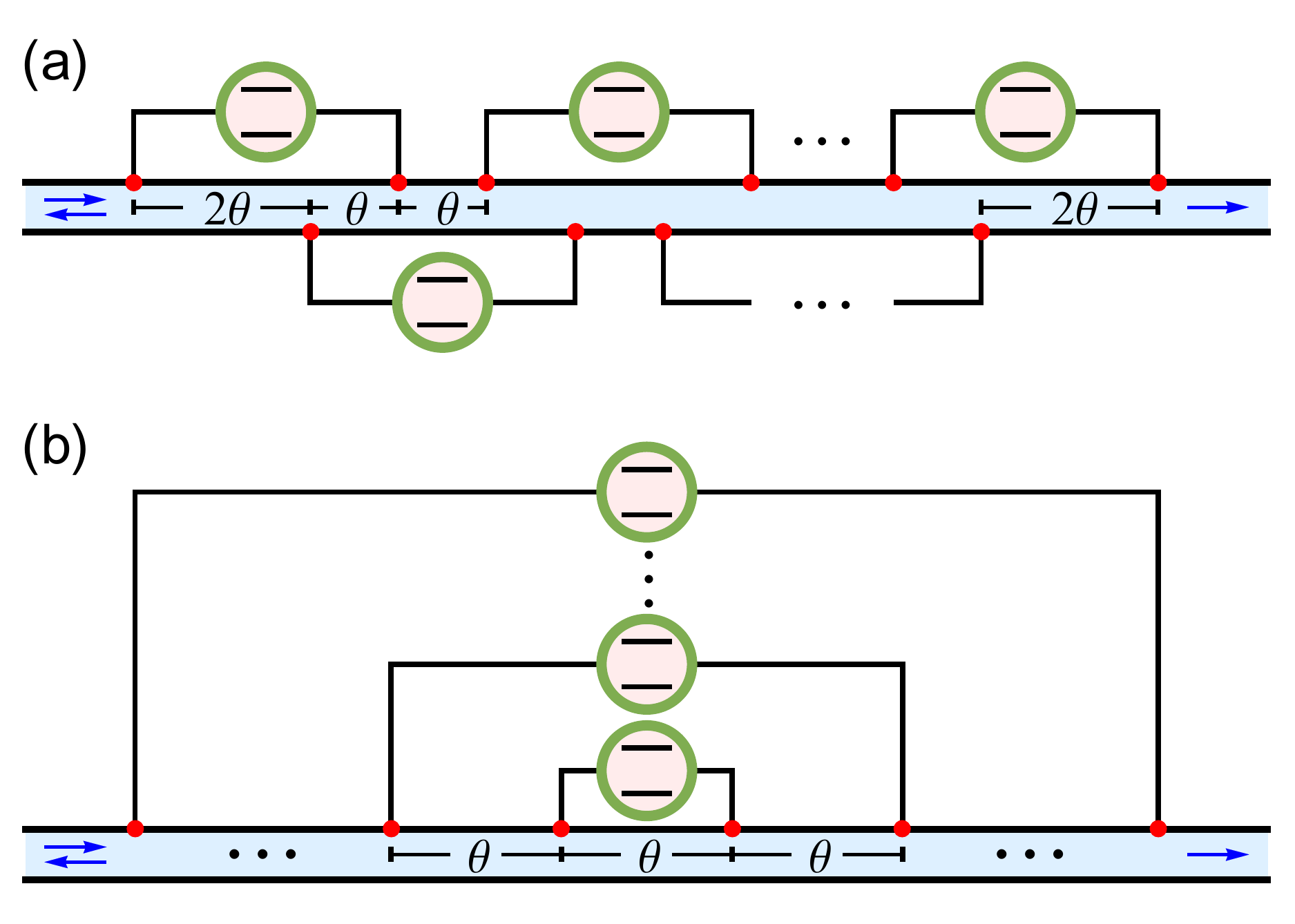}
	\caption{Sketches of 1D chains of (a) braided giant atoms, and (b) nested atoms.}
	\label{SketchBN}
\end{figure}
\begin{figure*}[t]
	\centering
	\includegraphics[width=\textwidth]{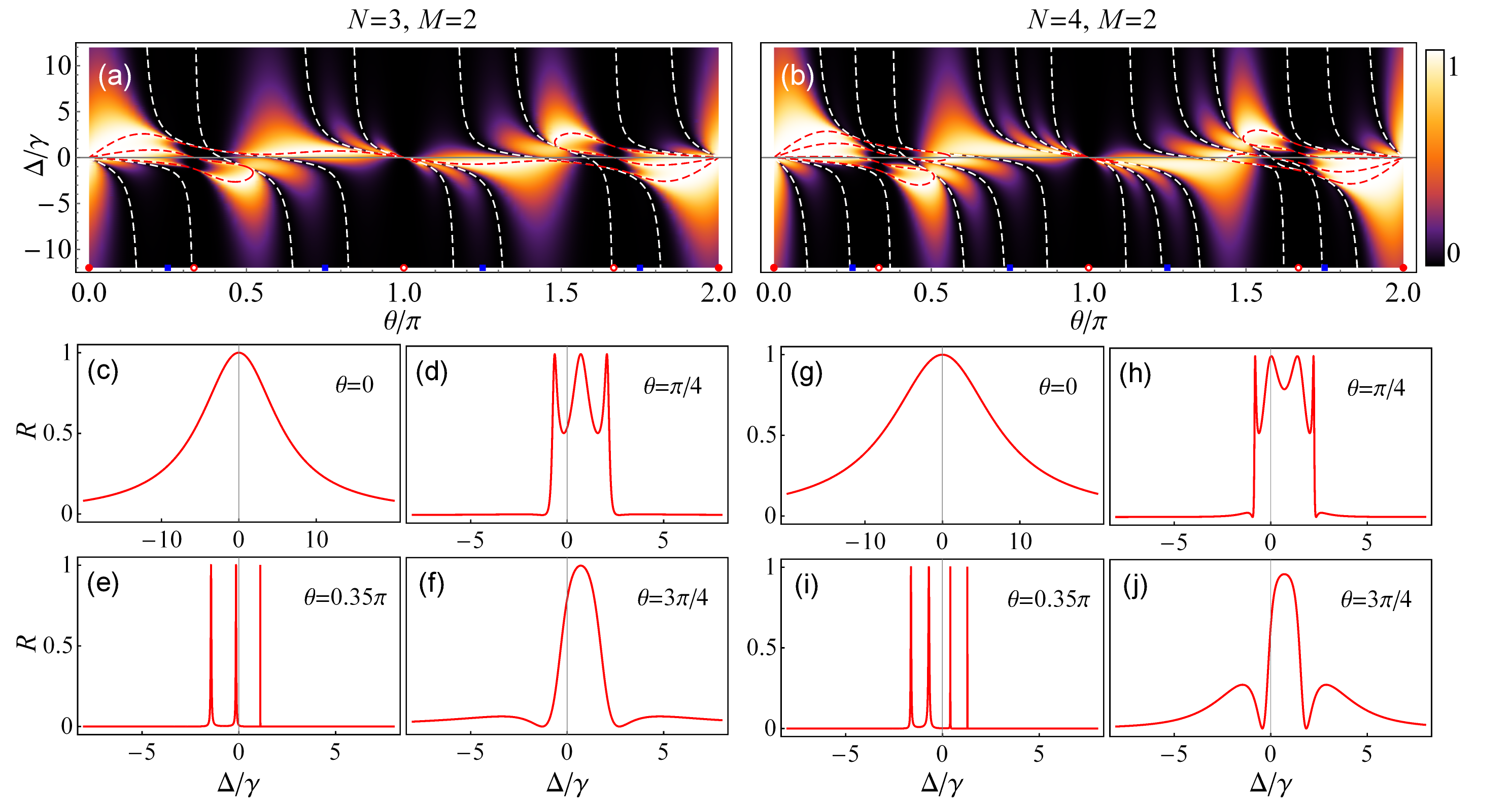}
	\caption{Reflectance $R$ for braided giant atoms shown in Fig.~\ref{SketchBN}(a) as functions of detuning $\Delta$
		and phase $\theta$, with (a) $N=3, M=2$, and (b) $N=4, M=2$. The red (white) dashed lines are used to mark the
		locations of the total (zero) reflection. Some special phase delays are indicated by the red circles (decoupling), 
		the red disks (superradiance), and the blue squares (symmetrical non-Lorentzian-spectrum), respectively. 
		The curves in panels (c)–(f) [(g)–(j)] show the cross sections of panel (a) [(b)] at phases $\theta=0$, 
		$\theta=\pi/4$, $\theta=0.35\pi$, and $\theta=3\pi/4$, respectively. 
		}
	\label{BraidedSpectrum}
\end{figure*}
\begin{figure*}[t]
	\centering
	\includegraphics[width=\textwidth]{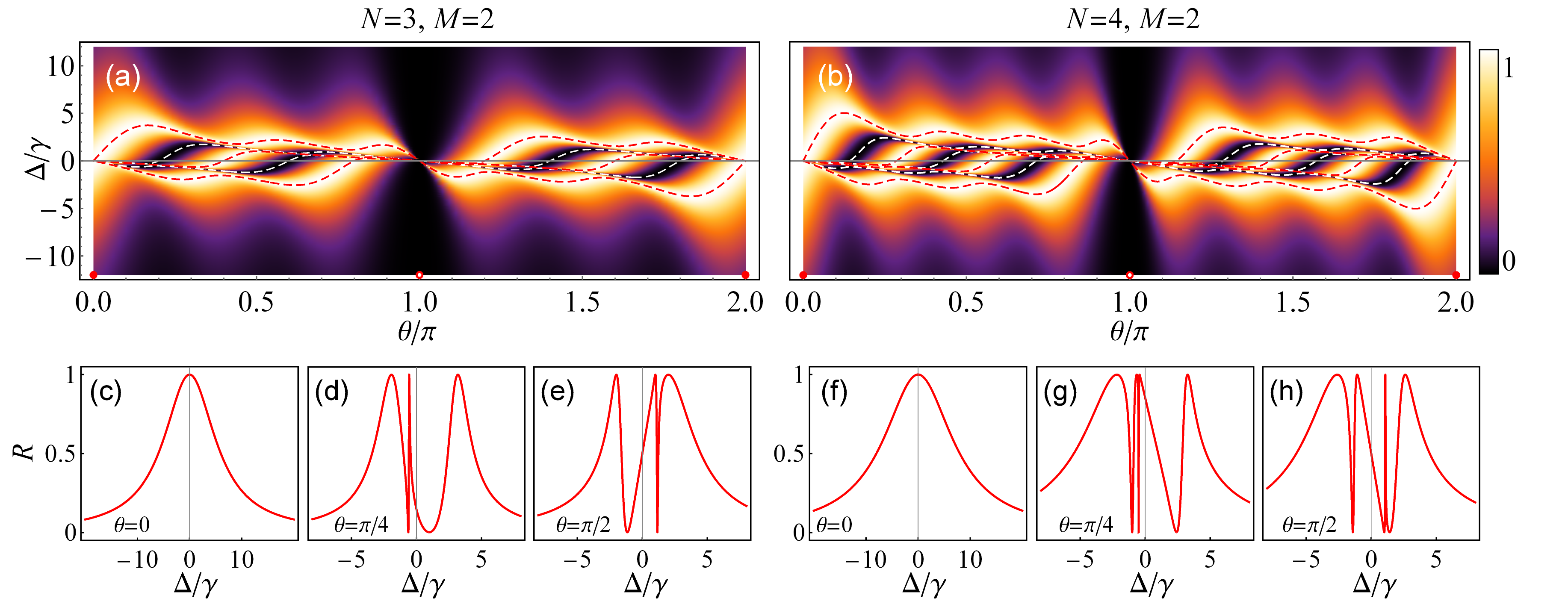}
	\caption{Reflectance $R$ for nested giant atoms shown in Fig.~\ref{SketchBN}(b) as functions of detuning $\Delta$
		and phase $\theta$, with (a) $N=3, M=2$, and (b) $N=4, M=2$. The red (white) dashed lines are used to mark the
		locations of the total (zero) reflection. Some special phase delays are indicated by the red circles (decoupling) and the red disks 
		(superradiance), respectively. The curves in panels (c)–(e) [(f)–(h)] 
		show the cross sections of panel (a) [(b)] at phases $\theta=0$, $\theta=\pi/4$, and $\theta=\pi/2$, 
		respectively.}
	\label{NestedSpectrum}
\end{figure*}
\section{\label{DerivationTM}Derivation of transfer matrix for an array of separate atoms}
Note that for an array of separate atoms, the coupling points at $x_{im}$ and $x_{i,m-1}$ are always adjacent to 
each other. Thus Eqs.~\eqref{Eot}-\eqref{Eof} become
\begin{subequations}
\begin{equation}
-\mathrm{i} v_{\mathrm{g}}(t_{im}-t_{i,m-1})e^{\mathrm{i}kx_{im}}+V_{im}f_{i}=0,
\label{cou1}
\end{equation}	
\begin{equation}
\mathrm{i} v_{\mathrm{g}}(r_{i,m+1}-r_{i,m})e^{-\mathrm{i}kx_{im}}+V_{im}f_{i}=0,
\label{cou2}
\end{equation}
\begin{eqnarray}
\Delta _{i}f_{i}-\frac{1}{2}\sum_{m=1}^{M_i}V_{im}\left[(t_{im}+t_{i,m-1})e^{\mathrm{i}kx_{im}}\right.
\nonumber
\\
\left.+(r_{i,m+1}+r_{im})e^{-\mathrm{i}kx_{im}}\right] =0.
\label{cou3}
\end{eqnarray}
\end{subequations}
Note that for the separate configuration, we have the following relations: $t_{i0}=t_{i-1,M_{i-1}}$, $r_{i,M_i+1}=r_{i+1,1}$.

Starting from Eqs.~\eqref{cou1} and \eqref{cou2}, after iterative calculation, we obtain the following relations
\begin{subequations}
	\begin{equation}
		t_{im}=t_{i-1,M_{i-1}}-\mathrm{i} \frac{1}{v_{\mathrm{g}}}\sum_{m'=1}^{m}V_{im'}e^{-\mathrm{i}kx_{im'}}f_{i},
		\label{tim}
	\end{equation}	
	\begin{equation}
r_{im}=r_{i+1,1}-\mathrm{i} \frac{1}{v_{\mathrm{g}}}\sum_{m'=m}^{M_i}V_{im'}e^{\mathrm{i}kx_{im'}}f_{i}.
\label{rim}
	\end{equation}
\end{subequations}
Substituting above results into Eqs.~\eqref{cou3} and using definition $\gamma_{im}=2V_{im}^2/v_{\mathrm{g}}$, we have
\begin{equation}
f_i=\frac{\sum_{m=1}^{M_i}V_{im}\left(e^{\mathrm{i}kx_{im}}t_{i-1,M_{i-1}}+e^{-\mathrm{i}kx_{im}}r_{i+1,1}\right)}{{\Delta _{i}}+\frac{\mathrm{i}}{2} \sum _{m,m'=1}^{M_i}\sqrt{\gamma _{im}\gamma _{im'}}e^{\mathrm{i}k|x_{im}-x_{im'}|}}.
\label{fi}
\end{equation}	
According to Eqs.~\eqref{tim}, \eqref{rim}, and \eqref{fi}, we can obtain Eq.~\eqref{recursiveM}, in which the scattering amplitudes on the left and right of the $i$th atom are connected by the transfer matrix.
\section{\label{Derivation2} Derivation of scattering coefficients for an array of periodically arranged identical giant atoms in a separated configuration}
It can be verified that the determinant of the matrix $\tilde{\mathbf T}$ is 1. Based on Abeles's theorem \cite{Abeles-AnnPhys1950}, we can write the matrix $\tilde{\mathbf{T}}^{N}$ in terms of Chebyshev polynomials of the second kind:
\begin{equation}
\tilde{\mathbf T}^{N}=U_{N-1}(y)\tilde{\mathbf T}-U_{N-2}(y)\mathbf{I},
	\label{decomposeTN}
\end{equation}
where 
\begin{subequations}
	\begin{equation}
U_{N-1}\left( y \right) =\left\{ \begin{array}{l}
	\frac{\sin N\Lambda}{\sin \Lambda},\Lambda =\arccos \left( y \right),~~|y|\le 1
	\\
	\\
	\left[ \mathrm{sgn} \left( y \right) \right] ^{N-1}\frac{\sin N\Lambda}{\sin \Lambda},
	\\
	\\
	\Lambda =\ln \left( |y|+\sqrt{|y|^2-1} \right),~~|y|>1
\end{array} \right.
		\label{Uexpression}
	\end{equation}	
	\begin{equation}
y=\frac{1}{2}\mathrm{Tr}(\tilde{\mathbf{T}})=\cos\phi+\xi\sin\phi,
		\label{y}
	\end{equation}
\end{subequations}
and $\mathbf{I}$ is the identity matrix. Chebyshev polynomial $U_{N-1}(y)$ has $N-1$ roots
\begin{equation}
y_{ s } = \cos\left(\frac{s\pi}{N}\right),
	\label{yroot}
\end{equation}
with $s=1,2,\cdots, N-1$. And the following identities will be used in this paper
\begin{subequations}
\begin{equation}
U_N(\pm1)=(\pm 1)^N(N+1),
\label{specialvalue1}
\end{equation}	
\begin{equation}
U^2_{N-1}(y)+U^2_{N-2}(y)-2yU_{N-1}(y)U_{N-2}(y)=1.
\label{idetityn1}
\end{equation}
\end{subequations}
Using Eqs.~\eqref{finalt}, \eqref{finalr}, \eqref{decomposeTN}, and \eqref{idetityn1}, we can obtain the explicit analytical expressions for scattering coefficients, as shown by Eqs.~\eqref{abst} and \eqref{absr} in the main text. 
\section{\label{SpectraBraided}Spectra for an array of braided giant atoms}
Here, we consider a 1D chain containing $N$ identical giant atoms with two connection points each. Each pair of neighboring atoms is in a braided configuration. The phase delays satisfy $\theta_{i,2}-\theta_{i+1,1}=\theta_{i+2,1}-\theta_{i,2}=\theta$, and all the bare decay rates are equal with $\gamma_{im}=\gamma$, as shown in Fig.~\ref{SketchBN}(a). 
In Figs.~\ref{BraidedSpectrum}(a) and \ref{BraidedSpectrum}(b), we plot the reflectance as functions of the detuning $\Delta$ and the phase delay $\theta$ for $N=3$ and $N=4$, respectively. 
The total reflection points with $R = 1$ are marked by the red dashed lines. And the reflection minima with $R=0$ due to destructive interference are marked by the white dashed lines. In addition, for a phase factor $\theta\in[0,\pi]$, we have relation $R(\Delta,\theta)=R(-\Delta,2\pi-\theta)$.
Thus, without loss of generality we show in Figs.~\ref{BraidedSpectrum}(c)-\ref{BraidedSpectrum}(f) [Figs.~\ref{BraidedSpectrum}(g)-\ref{BraidedSpectrum}(j)] the cross sections of Fig.~\ref{BraidedSpectrum}(a) [Fig.~\ref{BraidedSpectrum}(b)] at some typical phase delays in the region $\theta\in[0,\pi]$.
The detailed characteristics of the reflection spectra for different $\theta$ are summarized below.

\begin{itemize}
\item There are at most $N$ total reflection points and $N-1$ zero reflection points in the reflection spectrum. In some regions of $\theta$, the number of total reflection points decreases to one (zero) for $N\in\mathbb{O}^{+}$ ($N\in\mathbb{E}^{+}$), as shown in Figs.~\ref{BraidedSpectrum}(a) and \ref{BraidedSpectrum}(b). This is different from small atoms \cite{Tsoi-PRA2008,Mukhopadhyay-PRA2019} and separate giant atoms (see Sec.~\ref{SpectraSG}), where only one total reflection point appears if the atoms are not decoupled from the waveguide.

\item When $\theta=2n\pi$ [$n\in\mathbb{N}$, marked by the red disks in Figs.~\ref{BraidedSpectrum}(a) and \ref{BraidedSpectrum}(b)], one can obtain a Lorenzian spectrum that is symmetric at $\Delta=0$,
exhibiting a Dicke-type superradiant structure. Namely, the spectrum has a width that scales linearly as the size of the chain and equals $N\Gamma_{\mathrm{eff}}=4N\gamma$ (Note that each atom has an effective decay $\Gamma_{\mathrm{eff}}=4\gamma$ for this case.),
as shown in Figs.~\ref{BraidedSpectrum}(c) and \ref{BraidedSpectrum}(g) (with $\theta=0$).

\item When $\theta=(2n+1)\pi/4$ ($n\in\mathbb{N}$), one can obtain a non-Lorenzian spectrum structure that is symmetric at the Lamb shift. And there are $N-1$ ($N-2$) zero reflection points symmetrically distributed on both sides of the main peak for $N\in\mathbb{O}^{+}$ ($N\in\mathbb{E}^{+}$). 
In the range of $\theta\in [0,2 \pi]$, the points exhibiting this kind of spectrum structure are $\theta=\pi/4$, $\theta=3\pi/4$, $\theta=5\pi/4$, and $\theta=7\pi/4$, as indicated by the blue squares in Figs.~\ref{BraidedSpectrum}(a) and \ref{BraidedSpectrum}(b). The cross sections of Fig.~\ref{BraidedSpectrum}(a) [Fig.~\ref{BraidedSpectrum}(b)] at phase delays $\theta=\pi/4$ and $\theta=3\pi/4$ are shown in Figs.~\ref{BraidedSpectrum}(d) and \ref{BraidedSpectrum}(f) [Figs.~\ref{BraidedSpectrum}(h) and \ref{BraidedSpectrum}(j)]. 

\item When $\theta=(2n+1)\pi/3$ [$n\in\mathbb{N}$, indicated by the red circles in Figs.~\ref{BraidedSpectrum}(a) and \ref{BraidedSpectrum}(b)], the effective decay of each atom vanishes, and thus the atomic array decouples from the waveguide. 
Among these values of $\theta$, when $\theta=2m\pi\pm\pi/3$ ($m\in\mathbb{N}$), e.g., $\theta=\pi/3$ and $\theta=5\pi/3$ in the range of $\theta\in[0,2\pi]$, the so called decoherence-free interactions with strength $g_{i,i+1}=\pm{\sqrt{3}}\gamma/{2}$ exist between neighboring atoms. Whereas when $\theta=(2m+1)\pi$, e.g., $\theta=\pi$ in the range of $\theta\in[0,2\pi]$, both the effective decays and the coherent interactions vanish. In Figs.~\ref{BraidedSpectrum}(e) and \ref{BraidedSpectrum}(i), we plot the reflection spectra when the phase $\theta$ slightly deviates from the decoherence-free-interaction point $\pi/3$. In this case, one can ensure that $g_{i,i+1}\simeq\pm{\sqrt{3}}\gamma/{2}$, and at the same time the effective decay of each atom obtains a small value  $\Gamma_{\mathrm{eff},i}\ll g_{i,i+1}$. Thus the tight-binding atomic chain formed by the nearly decoherence-free interactions can interact with photons in the waveguide. The corresponding reflection spectrum describes the energy structure of the atomic chain. In the main text, we discuss a more interesting example of an SSH-type topological atomic chain resulting from the decoherence-free interactions (see Sec.~\ref{BraidedDFI}).  

\end{itemize}
\section{\label{SpectraNested}Spectra for an array of nested giant atoms}
In this section, we consider $N$ nested giant atoms with two connection points each. The phase delays between neighboring coupling points are equal with $\theta$, and all the bare decay rates are equal with $\gamma_{im}=\gamma$, as shown in Fig.~\ref{SketchBN}(b). 
In Figs.~\ref{NestedSpectrum}(a) and \ref{NestedSpectrum}(b), we plot the reflectance as functions of the detuning $\Delta$ and the phase delay $\theta$ for $N=3$ and $N=4$, respectively. For a phase factor $\theta\in[0,\pi]$, we have relation $R(\Delta,\theta)=R(-\Delta,2\pi-\theta)$.
The total reflection peaks and the zero reflection dips are marked by the red and white dashed lines, respectively. 
The detailed characteristics of the reflection spectra for different $\theta$ are summarized below.

\begin{itemize}
\item There are always $N$ total reflection points and $N-1$ zero reflection points in the reflection spectra, except for $\theta=n\pi$ ($n\in\mathbb{N}$). Each zero reflection point is located between two total reflection points, as shown by the red and white dashed lines in Figs.~\ref{NestedSpectrum}(a) and \ref{NestedSpectrum}(b). This feature is also shown by the cross sections [see Fig.~\ref{NestedSpectrum}(c)-\ref{NestedSpectrum}(h)].

\item When $\theta=2n\pi$ [$n\in\mathbb{N}$, marked by the red disks in Figs.~\ref{NestedSpectrum}(a) and \ref{NestedSpectrum}(b)], one can obtain a Lorenzian spectrum that is symmetric at $\Delta=0$. The width is $N\Gamma_{\mathrm{eff}}=4N\gamma$, exhibiting a Dicke-type superradiant structure (Note that each atom has an effective decay $\Gamma_{\mathrm{eff}}=4\gamma$ for this case.), as shown in Figs.~\ref{NestedSpectrum}(c) and \ref{NestedSpectrum}(f) (with $\theta=0$).

\item When $\theta=(2n+1)\pi$ ($n\in\mathbb{N}$), the effective decay of each atom vanishes, and the atomic array decouples from the waveguide. 
For example, the decoupling point is $\theta=\pi$ in the range of $\theta\in[0,2\pi]$, as indicated by the red circles in Figs.~\ref{NestedSpectrum}(a) and \ref{NestedSpectrum}(b). 
\end{itemize}
\bibliography{MS-PYP-Sep-23-2023}
\end{document}